\documentclass[12pt]{article}
\pdfoutput=1

\global\arraycolsep=1pt
\usepackage{graphicx}
\usepackage{amsmath}
\usepackage{amssymb}
\usepackage{amsthm}
\usepackage[export]{adjustbox}
\usepackage{hyperref}
\usepackage{microtype}
\usepackage{ytableau}

\newcommand{\CR}{\nonumber \\}
\newcommand{\gq}{\mathfrak{q}}
\newcommand{\gd}{\mathfrak{d}}
\makeatletter
 
 \@addtoreset{equation}{section}
\makeatother

\begin{document}

\title{\textbf{The MacMahon $R$-matrix}}

\author{{\bf Hidetoshi Awata$^a$}\footnote{awata@math.nagoya-u.ac.jp},
\ {\bf Hiroaki Kanno$^{a,b}$}\footnote{kanno@math.nagoya-u.ac.jp},
\ {\bf Andrei Mironov$^{c,d,e}$}\footnote{mironov@lpi.ru; mironov@itep.ru},\\
\ {\bf Alexei Morozov$^{d,e}$}\thanks{morozov@itep.ru},
\ {\bf Kazuma Suetake$^a$}\footnote{m14020z@math.nagoya-u.ac.jp},
\ \ and \ {\bf Yegor Zenkevich$^{d,f,g}$}\thanks{yegor.zenkevich@gmail.com}}
\date{}
\maketitle

\vspace{-8cm}

\begin{center}
\hfill FIAN/TD-18/18\\
\hfill IITP/TH-18/18\\
\hfill ITEP/TH-30/18
\end{center}

\vspace{4.3cm}

\begin{center}
$^a$ {\small {\it Graduate School of Mathematics, Nagoya University,
Nagoya, 464-8602, Japan}}\\
$^b$ {\small {\it KMI, Nagoya University,
Nagoya, 464-8602, Japan}}\\
$^c$ {\small {\it Lebedev Physics Institute, Moscow 119991, Russia}}\\
$^d$ {\small {\it ITEP, Moscow 117218, Russia}}\\
$^e$ {\small {\it Institute for Information Transmission Problems, Moscow 127994, Russia}}\\
$^f$ {\small {\it Dipartimento di Fisica, Universit\`a di Milano-Bicocca,
Piazza della Scienza 3, I-20126 Milano, Italy}}\\
$^g$ {\small {\it INFN, sezione di Milano-Bicocca,
I-20126 Milano, Italy
  }}
\end{center}

\vspace{.5cm}

\begin{abstract}
  We introduce an $R$-matrix acting on the tensor product of MacMahon
  representations of Ding-Iohara-Miki (DIM) algebra
  $U_{q,t}(\widehat{\widehat{\mathfrak{gl}}}_1)$. This $R$-matrix acts
  on pairs of $3d$ Young diagrams and retains the nice symmetry of the
  DIM algebra under the permutation of three deformation parameters
  $q$, $t^{-1}$ and $\frac{t}{q}$. We construct the intertwining
  operator for a tensor product of the horizontal Fock representation
  and the vertical MacMahon representation and show that the
  intertwiners are permuted using the MacMahon $R$-matrix.
\end{abstract}

\newpage
\tableofcontents

\section{Introduction}
\label{sec:introduction}
Ding-Iohara-Miki (DIM) algebra \cite{DI,Miki}, also denoted by
$U_{\gq,\gd}(\widehat{\widehat{\mathfrak{gl}}}_1)$, or
$U_{q,t}(\widehat{\widehat{\mathfrak{gl}}}_1)$ is a $t$-deformation of
the algebra of volume-preserving diffeomorphisms of the quantum torus \cite{GKV,VV},
which can also be described as $qW_{1+\infty}$-algebra, or a
deformation of the double loop algebra of $\mathfrak{gl}_1 =
\mathbb{C}^\times$ \cite{Moody}. The DIM algebra is an extremely elegant object. It has three
deformation parameters $q_1 = q =\gd \gq^{-1} $ (coming from the
quantization of the torus), $q_2 = t^{-1} = \gq^{-1} \gd^{-1}$ (the
$t$-deformation) and $q_3 = \frac{t}{q} = \frac{1}{q_1 q_2} = \gq^2$,
which are treated on equal footing. In fact, the algebra is invariant
with respect to the permutations of $q_i$. It also enjoys an
$SL(2,\mathbb{Z})$ symmetry\footnote{For possible extensions to $SL(3,\mathbb{Z})$ symmetry see \cite{LV,SS,MMZ,Zab}.}, corresponding to the mapping class group
of the quantum torus. There are two central charges in the algebra,
which form a two-dimensional representation of the $SL(2,\mathbb{Z})$
automorphism group.

A natural\footnote{Though, probably not the most general: in the
MacMahon representations one of the central charges is fixed to be unit (we use the multiplicative convention for levels), while
generally the both central charges can be chosen arbitrarily.} representation
of the DIM algebra seems to be the MacMahon module $\mathcal{M}(K;v)$, i.e.\ a
linear space spanned by \emph{plane partitions,} or $3d$ Young
diagrams. Here is an example of a $3d$ Young diagram:
\begin{equation}
  \label{eq:11}
\includegraphics[valign=c]{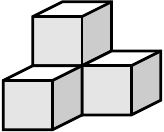} \quad = \quad [[2,1],[1]].
\end{equation}

We denote the $3d$ diagrams as sequences of ordinary Young diagrams lying at the successive horizontal layers,
as seen on the r.h.s.\ of Eq.~(\ref{eq:11}). Metaphorically, one can
think of an additional dimension in the diagrams as related to an
additional \emph{loop} in the definition of the algebra (compared,
e.g., to the affine or Virasoro algebras). The MacMahon representation has the
central charge vector $(1,K)$ (with $K$ arbitrary). Permutations of the
three deformation parameters $q_i$ act on this representation by
permutations of the \emph{axes} of the $3d$ Young diagram, i.e.\ by
$3d$ ``transpositions''.

There are other representations of the DIM algebra, which are in fact
\emph{reductions} of the MacMahon one. If one puts the central charge
$K$ to a particular quantized value $K=q_i$, then the MacMahon
representation becomes reducible. After factoring out the invariant
subspace, one gets the space of \emph{ordinary} Young diagrams. The
resulting representation is called the Fock (or free field)
representation $\mathcal{F}(v)$. More generally, when $K=q_1^a q_2^b
q_3^c,~a,b,c \in \mathbb{Z}$, the MacMahon representation reduces to a
representation spanned by plane partitions with a ``pit''~\cite{BFM},~\cite{FJMM}.
Further reduction is provided by the vector representation, in which
the central charges are trivial and the vectors can be thought of as
columns of boxes, i.e.\ $1d$ Young diagrams.

The DIM algebra is a (quasi triangular) Hopf algebra and has a
universal $R$-matrix.  In \cite{Awata:2016mxc, Awata:2016bdm}, we have
explicitly constructed the $R$-matrix acting on the product of Fock
representations and elucidated its role in refined topological strings
and the Nekrasov partition function\footnote{Similar $R$-matrix was
  considered from the point of view of geometric representation theory
  in~\cite{Smirnov:2013hh}.}. In the present work, we follow the same
program and construct the DIM $R$-matrix for the MacMahon
representations.

For the Fock spaces, there is a well-known construction \cite{Awata:2011ce} of
intertwining operators, which turn out to be \emph{the refined
  topological vertices} $C_{\lambda \mu \nu}(q,t)$ of topological
string theory \cite{IKV,GIKV,AK0,Taki,AK}. The labels $\lambda$, $\mu$ and $\nu$ in the vertex are
$2d$ Young diagrams denoting the vectors in the three Fock spaces being
intertwined. Here we are going to generalize the intertwiner
construction so that one of the labels in the vertex is the $3d$ Young
diagram. This indicates the existence of a generalization of refined
topological string theory, in which the $3d$ Young diagrams live on the
edges along the preferred direction of the toric diagram.

The parameters $q$ and $t$ are perhaps most familiar from the theory of
Macdonald polynomials \cite{MacD}. The basis of Macdonald polynomials in fact
plays a distinguished role in the Fock representation of DIM
algebra. The $R$-matrix is then conveniently written as a set of
matrix elements in the Macdonald basis. Our construction for the
$R$-matrix in the MacMahon representations is similar, but the role of
Macdonald polynomials is played by \emph{triple} Macdonald
polynomials \cite{Zenk}.

More concretely, in this note, we try to construct the intertwining
operator for the MacMahon representation as the vertical
representation.  In \cite{Awata:2011ce}, the authors chose the Fock
representation and constructed the intertwiner that reproduced the
refined topological vertex. We want to replace the Fock representation
with the MacMahon representation.  One of the features of the MacMahon
representation is that it has a continuous value $K$ for the second
level $\gamma_2$ (see  sec.\ref{sec:def} for the notation). Consequently, we also need vertex operator
representations that allow $\gamma_2$ to be continuous.  At the
best of our knowledge, such a vertex operator representation is not
known at the moment, though a continuous shift of $\gamma_2$ may
be achieved by some judicious choice of the zero mode sector.  To
overcome this problem, we make use of the idea of constructing the
MacMahon representation from an infinite tensor product of Fock
representations \cite{FJMM}.  A similar idea is employed for
constructing the Fock representation from the vector representation
\cite{FFJMM}.  Therefore, in order to support the validity of our approach, we
first start with the intertwiner for the vector representation and
construct the Fock intertwiner in a way parallel to \cite{FFJMM} where
the Fock representation was constructed from a certain infinite tensor
product of vector representations.  It turns out that, with this method, we can obtain
the same Fock intertwiner as \cite{Awata:2011ce} did and, as a by-product,
also an additional insight into relevant horizontal representations. Thus, we believe we can lift the construction in
\cite{FFJMM,FJMM} to the level of intertwiners.

In the remaining part of the introduction, as a warm-up, we offer a
construction of the intertwiner and of the $R$-matrix for the vector
representations in order to prepare for
technicalities of the main text. The remaining part of the paper is organized as
follows. We introduce the definition of the DIM algebra in
sec.\ref{sec:def} and present the known representations of the algebra
in secs.~\ref{sec:ver}--\ref{sec:hor}. In particular, we review the
idea of taking an infinite tensor product following
\cite{FFJMM} and \cite{FJMM} in  sec.\ref{sec:ver}.  The crucial
point here is how to regularize the infinite products.
In sec.~\ref{sec:dim-r-matrix},  based on the formula of the diagonal
part of the universal $R$-matrix \cite{FJMM2}, we compute the
$R$-matrix for the MacMahon representations.
We then construct the intertwining operator for the MacMahon representations
in a way parallel to \cite{FFJMM, FJMM} in sec.~\ref{sec:int}.
In sec.~\ref{sec:r-matrix-from}, we confirm that
the commutator of such intertwiners indeed gives the same MacMahon $R$-matrix
computed in sec.~\ref{sec:dim-r-matrix}.
We present our conclusions in sec.~\ref{sec:conclusions}.

\subsection{$qW_{1+\infty}$ algebra and the vector representation}
\label{sec:qw_1+-lgebra-vect}
In this part of the introduction, we deal with a ``toy model'' of the
representations and intertwiners, which we are going to study in the
main part of the paper.

\paragraph{The toy algebra.}
\label{sec:toy-algebra}
Instead of the DIM algebra, we take its $t=1$ limit: $qW_{1+\infty}$ algebra,
which is a Lie algebra spanned by the
generators $W_{m,n}$ satisfying the relations
\begin{equation}
  \label{eq:23}
  [W_{m_1,n_1}, W_{m_2,n_2}] = (q^{n_1 m_2} - q^{n_2 m_1})
  W_{m_1+m_2,n_1+n_1} + (c_1 n_1 + c_2 m_1) q^{\frac{n_1 m_1}{2}} \delta_{m_1+m_2,0}
  \delta_{n_1 + n_2,0},
\end{equation}
where $c_{1,2}$ are the central charges. There are two grading
operators $d_1$ and $d_2$ counting $m$ and $n$ in $W_{m,n}$. The
relations are manifestly $SL(2,\mathbb{Z})$-invariant, if we assume
that the labels $(m,n)$ and the central charges $(c_1,c_2)$ transform
as $2d$ vectors under $SL(2,\mathbb{Z})$-transformations.

\paragraph{The toy representation.}
\label{sec:toy-representation}
We further simplify our toy model by putting $c_{1,2}
=0$. Then the algebra~(\ref{eq:23}) can be represented by difference
operators acting on functions of one variable $z$:
\begin{equation}
  \label{eq:24}
  \rho(W_{m,n}) = z^m q^{n z\partial_z}.
\end{equation}
We call representation $\rho$ the vector representation. This is
nothing but the \emph{evaluation representation} of the
algebra~(\ref{eq:23}). As usual, there are two viewpoints on the
evaluation representation. In the first convention, the representation
is seen as a \emph{finite}-dimensional (in our case one-dimensional)
space of functions $f(z)$, while, within the second approach, the
representation is an \emph{infinite}-dimensional space of vectors, in
which the parameter $z$ is not a variable, but is fixed.

In the case under discussion, one can also use the equivalence of two
conventions and say that the states in the vector representation,
instead of being functions of $z$, are now labelled by an integer and
a complex parameter, $|z,i\rangle$, so that
\begin{equation}
  \label{eq:25}
  \rho(W_{m,n})|z,\ell \rangle = z^m | z, \ell +n\rangle.
\end{equation}
The parameter $z$ is fixed in this definition and is called the
spectral parameter of the vector representation. Evidently, this is
just a rewriting of Eq.~(\ref{eq:24}) with $|z, \ell \rangle$ corresponding
to $f(z q^\ell)$.

\paragraph{The toy intertwiner.}
\label{sec:toy-intertwiner}
The toy algebra~(\ref{eq:23}) is a Hopf algebra with the standard
Lie-algebraic coproduct
\begin{equation}
  \label{eq:26}
  \Delta(W_{m,n}) = W_{m,n} \otimes 1 + 1 \otimes W_{m,n}.
\end{equation}
One can then build tensor products of representations. In particular, one
can take the tensor product of the Fock representation
of~(\ref{eq:23}), which has $c_1 = \log_{q_3} K_1 = 1, c_2=
\log_{q_3} K_2 = 0$, and the vector representation.

There is an intertwining operator
$\Psi$, which acts from this tensor product into a single Fock
representation. It is just the standard ($q$-deformed) vertex operator
\begin{equation}
  \label{eq:27}
  \Psi |z, \ell \rangle \otimes \, \cdots = \exp \left( \sum_{n \geq 1}
    \frac{1}{n} z^n q^{\ell n} a_{-n} \right) \exp \left( -\sum_{n \geq 1}
    \frac{1}{n} z^{-n} q^{-\ell n} a_n \right).
\end{equation}
The intertwiners commute up to contact terms, which is the reflection
of the fact that the coproduct~(\ref{eq:26}) is symmetric, and thus
the $R$-matrix of the algebra~(\ref{eq:23}) is trivial.

\paragraph{A small generalization and the toy $R$-matrix.}
\label{sec:small-gener-toy}
Leaving the rigorous derivation for sec.~\ref{sec:int}, we give here
the vector intertwining operator for $t \neq q$ (see also \cite{FHHSY}):
\begin{equation}
  \label{eq:6}
  \Psi ( |z, \ell \rangle \otimes \, \cdot) = \exp \left( \sum_{n \geq 1}
    \frac{1}{n} \frac{1-t^n}{1-q^n} t^{-n} z^n q^{\ell n} a_{-n} \right) \exp \left( -\sum_{n \geq 1}
    \frac{1}{n} q ^nz^{-n} q^{-\ell n} a_n \right).
\end{equation}
The commutation of intertwiners~(\ref{eq:6}) yields a
\emph{nontrivial} (though diagonal) $R$-matrix:
\begin{equation}
  \label{eq:8}
  \Psi ( |z,k\rangle \otimes \Psi ( |w,\ell \rangle \otimes \, \cdot)) =
  R_{k\ell}\left( \frac{z}{w} \right)   \Psi ( |w,\ell\rangle \otimes \Psi ( |z,k\rangle \otimes \, \cdot)),
\end{equation}
where
\begin{equation}
  \label{eq:19}
  R_{k\ell}\left( \frac{z}{w} \right) = \frac{\left( \frac{z}{w} q^{k-\ell}
      t;q \right)_{\infty} \left( \frac{z}{w} q^{k-\ell} \frac{q}{t} ;q
    \right)_{\infty}}{\left( \frac{z}{w} q^{k-\ell} ;q \right)_{\infty} \left( \frac{z}{w} q^{k-\ell} q;q \right)_{\infty}}.
\end{equation}
This $R$-matrix, though may seem simple, actually encodes in
itself the information about all the representations of the DIM
algebra. This is because all the representations can be built from the
vector representation by taking appropriate tensor products so that
the intertwiners also become products of the basic
intertwiner~(\ref{eq:6}). This will be precisely our strategy in the
main part of the paper.


\section{Representations of DIM algebra}
\label{sec:intr-dim-algebra}

\subsection{Definition of the quantum toroidal algebra $U_{\gq,\gd}(\widehat{\widehat{\mathfrak{gl}}}_1)$}
\label{sec:def}

The quantum toroidal algebra $U_{\gq,\gd}(\widehat{\widehat{\mathfrak{gl}}}_1)$ has two deformation parameters $\gq, \gd$.
We introduce the structure function
\begin{align}
  g(z,w) &= (z-q_1w)(z-q_2w)(z-q_3w)
\end{align}
where our convention is
\begin{align}
  q_1 = \gd \gq^{-1}, \quad  q_2 = \gd^{-1} \gq^{-1}, \quad  q_3 = \gq^2, \qquad q_1q_2q_3 =1.
\end{align}
In this paper, we assume that $q_i$ are generic, that is, $q_1^a q_2^b q_3^c=1$ for integer $a$, $b$, $c$ if and only if $a=b=c$.
We also use
\begin{align}
  G(z) &= -\frac{g(z,1)}{g(1,z)}, \\
  G(w/z) &= \frac{(q_1-w/z)(q_2-w/z)(q_3-w/z)}{(1-q_1w/z)(1-q_2w/z)(1-q_3w/z)} = -\frac{g(w,z)}{g(z,w)} = G(z/w)^{-1}.
\end{align}

The generators of $U_{\gq,\gd}(\widehat{\widehat{\mathfrak{gl}}}_1)$ are
$E_k, F_k, K^{\pm1}, H_r$ ($k\in\mathbb{Z}, r\in\mathbb{Z}\setminus \{ 0\}$) and the central element $C$.
It is convenient to introduce the generating functions (currents)
\begin{align}
  E(z) = \sum_{k \in \mathbb{Z}} E_{k} z^{-k},  \quad
  F(z) = \sum_{k \in \mathbb{Z}} F_{k} z^{-k}, \quad
  K^{\pm}(z) = K^{\pm 1} \exp \left( \pm ( \gq - \gq^{-1} ) \sum_{r=1}^\infty H_{\pm r} z^{\mp r}
\right).
\end{align}
Note that $K^{\pm}(z)$ is expanded in negative (positive) powers of $z$.
With these currents, $U_{\gq,\gd}(\widehat{\widehat{\mathfrak{gl}}}_1)$ is defined
by the following relations\footnote{They also satisfy the Serre relations,
which we do not use in this paper, \cite{Miki,Tsymb}.}:
\begin{align}
  K^{\pm}(z) K^{\pm}(w) &= K^{\pm}(w) K^{\pm}(z), \label{KK1} \\
  \frac{g(C^{-1}z,w)}{g(Cz,w)} ~K^{-}(z) K^{+}(w) &= \frac{g(w,C^{-1}z)}{g(w,Cz)} ~K^{+}(w) K^{-}(z) \label{KK2}, \\
  g(z,w) ~K^{\pm} (C^{(1\mp 1)/2} z) E(w) &+ g(w,z) ~E(w) K^{\pm} (C^{(1\mp 1)/2} z) = 0, \\
  g(w,z) ~K^{\pm} (C^{(1\pm 1)/2} z) F(w) &+ g(z,w) ~F(w) K^{\pm} (C^{(1\pm 1)/2} z) = 0, \\
  \left[ E(z), F(w) \right] = \tilde{g} &\left( \delta (C \frac{w}{z}) ~K^{+}(z)  -  \delta (C \frac{z}{w}) ~K^{-}(w) \right), \label{EFcoef} \\
  g(z,w) ~E(z) E(w) &+ g(w,z) ~E(w) E(z) = 0, \\
  g(w,z) ~F(z) F(w) &+ g(z,w) ~F(w) F(z) =0.
\end{align}
 The coefficient $\tilde{g}$ in \eqref{EFcoef} affects only the relative normalization of currents $E(z), F(z)$ and $K^\pm(z)$.
In this paper, we choose
\begin{align}
  \tilde{g} = \frac{(1-q_1)(1-q_2)}{1-q_3^{-1}}.
\end{align}
There are essentially\footnote{We impose $K^- = (K^+)^{-1}$.} two central elements $C$ and $K^- = (K^+)^{-1}$,
and we define that the representation has level $(\gamma_1, \gamma_2)$ if $C=\gamma_1$ and $K^-=\gamma_2$.
We call representations with $C=1$ and $C=\gq$ vertical and horizontal representation, respectively\footnote{They correspond to the two choices of the Borel subalgebra,  ``vertical'' and ``horizontal'' are related by the \emph{spectral
    duality} automorphism $\mathcal{S}$ \cite{Miki,specdu1,specdu2,specdu3,specdu4,specdu5,Sham1,Sham2,Sham3} (a proof is
  given in Lemma A.5 of~\cite{FJMM2}), which is one of the generators of the $SL(2,\mathbb{Z})$
automorphism group of the DIM algebra.}.
The commutation relation of modes $H_r$ are read from \eqref{KK2},
\begin{align}
  [H_r, H_s]
  &= \delta_{r+s,0} \frac{[r]}{r} (\gq^r+\gq^{-r}-\gd^r-\gd^{-r}) \frac{C^r-C^{-r}}{\gq-\gq^{-1}} \CR
  &= \delta_{r+s,0} \frac{[r]}{r} \gq^r(1-q_1^r)(1-q_2^r) \frac{C^{r}-C^{-r}}{\gq-\gq^{-1}},
\end{align}
where we define $\gq$-integer by
\begin{equation}
  [r] := \frac{\gq^r-\gq^{-r}}{\gq-\gq^{-1}}.
\end{equation}
Note that in the vertical representations $H_r$ are mutually commuting.

Actually the vertex operator representation (see section \ref{sec:hor}),
which is the only horizontal representation to be considered
in this paper, satisfies stronger relations:
\begin{align}
  K^{\pm}(z) K^{\pm}(w) &= K^{\pm}(w) K^{\pm}(z), \\
  K^{+}(z) K^{-}(w) &= G(C^{-1}w/z) G(Cw/z)^{-1} ~K^{-}(w) K^{+}(z), \\
  K^{+} (z) E(w) &=  G(w/z) ~E(w) K^{+} (z), \\
  E(z) K^{-} (C w) &=  G(w/z) ~K^{-} (C w) E(z), \\
  K^{+} (C z) F(w) &= G(w/z)^{-1} ~F(w) K^{+} (C z), \\
  F(z) K^{-} (w) &= G(w/z)^{-1} ~K^{-} (w) F(z), \\
  \left[ E(z), F(w) \right] = \tilde{g} &\left( \delta (C \frac{w}{z}) ~K^{+}(z)  -  \delta (C \frac{z}{w}) ~K^{-}(w) \right), \\
  g(z,w) ~E(z) E(w) &+ g(w,z) ~E(w) E(z) = 0, \\
  g(w,z) ~F(z) F(w) &+ g(z,w) ~F(w) F(z) =0.
\end{align}

To define the tensor product of two representations, we need a coproduct of
$U_{\gq,\gd}(\widehat{\widehat{\mathfrak{gl}}}_1)$.
We use the following coproduct in this paper:
\begin{align}
  \Delta(E(z)) &= E(z) \otimes 1 + K^{-} (C_1z) \otimes E(C_1z),  \label{cpE} \\
  \Delta(F(z)) &=  F(C_2 z) \otimes K^{+}(C_2z)  +  1 \otimes F(z), \label{cpF} \\
  \Delta(K^{+} (z)) &= K^{+}(z) \otimes K^{+}(C_1^{-1} z), \label{cpK+} \\
  \Delta(K^{-} (z)) &= K^{-}(C_2^{-1} z) \otimes K^{-}(z), \label{cpK-} \\
  \Delta(C) &= C \otimes C,
\end{align}
where $C_1 = C \otimes 1$ and $C_2 = 1 \otimes C$.

We also have two grading operators $d_1$ and $d_2$ such that
\begin{gather}
  \label{eq:22}
  [d_1 , E(z)] = -E(z),\qquad [d_1, F(z)] = F(z),\qquad [d_1, H(z)] = 0,\\
  [d_2, E(z)] = z \partial_z E(z),\qquad   [d_2, F(z)] = z \partial_z F(z),
  \qquad   [d_2, H(z)] = z \partial_z H(z).
\end{gather}

\subsection{Vertical representations}
\label{sec:ver}

Since $C=1$ for the vertical representations,
the modes $H_r$ ($r\in\mathbb{Z}\setminus\{ 0\}$) commute with each other due to the relation \eqref{KK2}.
Then we can find a basis which simultaneously diagonalizes $H_r$.
In \cite{FFJMM, FJMM}, three kinds of vertical representations:
the vector, Fock and MacMahon representations have been defined.
In these representations, the basis on which $H_r$ acts diagonally is roughly speaking
labeled by 1D, 2D and 3D Young diagrams, respectively.
Accordingly, we can define the Fock representation as an irreducible subrepresentation of an appropriate infinite tensor product of
the vector representations \cite{FFJMM}. In a similar manner, we can construct the MacMahon representation from the Fock
representations \cite{FJMM}. The second level of the vector representation is $\gamma_2 =1$. However,
the regularization procedure required in the above procedure makes $\gamma_2$ non-trivial.
Consequently, the Fock representation has a quantized level $\gamma_2 = \gq$.
Moreover, quite interestingly the MacMahon representation allows continuous level $\gamma_2 =K$.
We can find a natural regularization for the Fock representation, while that for the MacMahon representation somehow looks
ambiguous and leads to an arbitrary value $K$, which we can interpret formally as a limit of $\gq^N$ ($N \to \infty$).
In summary, the explicit actions of the vector, Fock and MacMahon representations are
given by \eqref{vecKp}--\eqref{vecF}, \eqref{FockKp}--\eqref{FockF} and \eqref{MacK}--\eqref{MacF}, respectively.


\subsubsection{Vector representation}
The vector representation $V(v)$ consists of the basis $[v]_i$ labeled by $i \in \mathbb{Z}$ and depends on the spectral parameter $v$.
The action of the algebra is defined as
\begin{align}
  K^+(z) [v]_i &= \tilde{\psi}(q_1^i v/z) [v]_i, \label{vecKp} \\
  K^-(z) [v]_i &= \tilde{\psi}(q_1^{-i-1} z/v) [v]_i, \label{vecKm} \\
  E(z) [v]_i &= \mathcal{E} ~\delta(q_1^{i+1} v/z) [v]_{i+1}, \label{vecE} \\
  F(z) [v]_{i+1} &= \mathcal{F} ~\delta(q_1^{i+1} v/z) [v]_{i}, \label{vecF}
\end{align}
where the generating function of eigenvalues of $K^{\pm}(z)$  is expressed by
\begin{align}
  \tilde{\psi}(z) = \frac{(1-q_2^{-1}z)(1-q_3^{-1}z)}{(1-z)(1-q_1 z)} = \tilde{\psi}(q_1^{-1}/z).  \label{psitilde}
\end{align}
The multiplication factors $\mathcal{E}$ and $\mathcal{F}$ are determined from the choice of $\tilde{g}$:
\begin{align}
  \mathcal{E} \cdot \mathcal{F} = \tilde{g} \frac{(1-q_2^{-1})(1-q_3^{-1})}{1-q_1} = (1-q_2)(1-q_2^{-1}),
\end{align}
and, in this paper, we choose
\begin{align}
  \mathcal{E} = 1-q_2, \quad \mathcal{F} = 1-q_2^{-1}.
\end{align}
Since $\tilde{\psi}(0)=1$, the vector representation has the trivial level $(1,1)$.
It is also convenient to introduce a more fundamental function $\psi(z)$ such that
\begin{align}
  \psi(z) = \gq \frac{1-q_3^{-1}z}{1-z} = \psi(q_3/z)^{-1}, \quad
  \tilde{\psi}(z) = \psi(z) \psi(q_2^{-1}z)^{-1}.
\end{align}

\subsubsection{Fock representation}
\label{sec:Fockrep}

At first following \cite{FFJMM}, we outline the idea to obtain the Fock representation $(\rho, \mathcal{F}(v))$.
One can construct a natural tensor product of the vector representations
by making use of the coproduct of $U_{\gq,\gd}(\widehat{\widehat{\mathfrak{gl}}}_1)$.
We want to define an infinite tensor product
of the vector representations and find an irreducible subrepresentation,
whose basis is composed of partitions, that is, the Fock representation.
Let us consider the following tensor product with $q_2$-shifted spectral parameters:
\begin{align}
  \bigotimes_{i=1}^{N} V(q_2^{i-1}v)
  \ni \vert \lambda) =\bigotimes_{i=1}^{N} [q_2^{i-1}v]_{\lambda_i-1}, \quad
  \lambda = (\lambda_1, \ldots, \lambda_N) \in \mathbb{Z}^N.
\end{align}
Since $C=1$ for the vector representations, the coproduct \eqref{cpK+} and \eqref{cpK-} give
\begin{align}
\Delta^{N-1} (K^{\pm} (z)) = \overbrace{K^{\pm} (z) \otimes \cdots \otimes K^{\pm}(z)}^{N} \label{DNK}.
\end{align}
Similarly, the coproduct \eqref{cpE} and \eqref{cpF} implies
\begin{align}
\Delta^{N-1} (E(z)) &= \sum_{k=1}^N \overbrace{K^{-}(z) \otimes \cdots K^{-}(z)}^{k-1}
 \otimes E(z) \otimes \overbrace{1 \otimes \cdots \otimes 1}^{N-k}, \label{DNE} \\
\Delta^{N-1} (F(z)) &= \sum_{k=1}^N \overbrace{1 \otimes \cdots \otimes 1}^{k-1}
 \otimes F(z) \otimes \overbrace{K^{+}(z) \otimes \cdots \otimes K^{+}(z)}^{N-k} \label{DNF}.
\end{align}
The tensor product representation is defined by
$\rho_N (X(z)) := \rho^{\mathrm v}_{u_1} \otimes \cdots \otimes \rho^{\mathrm v}_{u_N} (\Delta^{N-1}(X(z)))$, where
$\rho^{\mathrm v}_{u_i}$ denotes the vector representation with the spectral parameter $u_i = q_2^{i-1}v$.

One can naturally view $\lambda \in \mathbb{Z}^N$ as $\lambda \in \mathbb{Z}^{N+1}$ with $\lambda_{N+1}=0$.
However $(\rho_N, \bigotimes_{i=1}^{N} V(q_2^{i-1}v))$ does not form an inductive system
because actions of $\rho_N$ and $\rho_{N+1}$ on $\vert \lambda )$ ($\lambda \in \mathbb{Z}^N$) are different.
For this reason, one cannot take a limit $N \to \infty$ naively.
In order to settle this problem, we modify the action of $(\rho_N, \bigotimes_{i=1}^{N} V(q_2^{i-1}v))$
to some $(\bar{\rho}_N, \bigotimes_{i=1}^{N} V(q_2^{i-1}v))$ so as to form an inductive system of
$U_{\gq,\gd}(\widehat{\widehat{\mathfrak{gl}}}_1)$-modules.
Then one can take the inductive limit $\rho = \displaystyle{\lim_{N\to\infty}}\bar{\rho}_N$
to find the Fock representation $(\rho, \mathcal{F}(v))$ as an irreducible subrepresentation of $\rho$:
\begin{align}
  \bigotimes_{i=1}^{\infty} V(q_2^{i-1}v)
  \supset \mathcal{F}(v)
  \ni \vert \lambda) =\bigotimes_{i=1}^{\infty} [q_2^{i-1}v]_{\lambda_i-1},
\end{align}
where $\lambda = (\lambda_1, \ldots, \lambda_{\ell(\lambda)}) \in \mathcal{P}: \mbox{set of partitions}$
and $\lambda_n=0 \ (n>\ell(\lambda) )$.

In the following, we describe the idea of modification in detail.
Actually, we have to modify only $\rho_N(K^\pm(z))$ and $\rho_N(F(z))$ keeping $\rho_N(E(z))$ intact.
Let us modify the action of $\rho_N$ to $\bar{\rho}_N$ so as to obtain the condition, for $X = K^\pm, E$ or $F$,
\begin{align}
  \bar{\rho}_N(X(z)) = \bar{\rho}_{N+M}(X(z)) \ \mbox{on} \ \vert \lambda), \ \lambda \in \mathbb{Z}^{N-1}, \ \quad \forall M \in \mathbb{N} \label{idea}
\end{align}
Then we can define the action of $\rho(X(z))$ on $\vert \lambda)$ ($\lambda \in \mathbb{Z}^{N-1}$) as $\bar{\rho}_N(X(z))$.
Therefore, we should search for a modified action which satisfies the condition \eqref{idea}.
At first, let us see that the action $\bar{\rho}_N(E(z))$ can be  the same as $\rho_N(E(z))$.
This action satisfies the condition \eqref{idea} due to the vanishing property,
\begin{align}
  &(K^-(z) \otimes E(z)) ([q_2^{N-1}v]_{-1} \otimes [q_2^{N}v]_{-1}) \\
  &= \tilde{\psi}(q_2^{-N+1}z/v) \delta(q_2^{N}v/z) [q_2^{N-1}v]_{-1} \otimes [q_2^{N}v]_{-1}
  =0.
\end{align}
Further, let us focus on the action of $K^\pm(z)$.
This time the actions of $\rho_N$ and $\rho_{N+M}$ differ.
We denote $K_N^\pm(z) = \bar{\rho}_N(K^\pm(z)) = \rho_N(K^\pm(z)) \times \beta_N^\pm = \Delta^{N-1}(K^\pm(z)) \times \beta_N^\pm$, where $\beta_N^\pm = \beta_N^\pm((v/z)^{\pm1})$ is a modification factor that satisfies $\beta_N^+(v/z) = \beta_N^-(z/v)$ as a rational function.
Since
\begin{align}
  K^+(z) [q_2^{N}v]_{-1} = \tilde{\psi}(q_1^{-1}q_2^{N}v/z) [q_2^{N}v]_{-1},
\end{align}
one gets a recursion relation for the modification factor $\beta_N^+$
\begin{align}
  1=&\frac{(\lambda \vert K_{N+1}^+(z) \vert \lambda)}{(\lambda \vert K_{N}^+(z) \vert \lambda)}
  = \frac{\beta^+_{N+1}}{\beta^+_N} ~\tilde{\psi}(q_1^{-1}q_2^{N}v/z).
\end{align}
This is equivalent to
\begin{align}
  \beta^+_N
  &= \tilde{\psi}(q_1^{-1} q_2^{N-1}v/z)^{-1} \beta^+_{N-1}
  = \psi(q_1^{-1}q_2^{N-1}v/z)^{-1} \psi(q_1^{-1}q_2^{N-2}v/z) \beta^+_{N-1} \CR
  &= \psi(q_1^{-1}q_2^{N-1}v/z)^{-1} \psi(q_1^{-1}v/z) \beta^+_1,
\end{align}
which determines $\beta^\pm_N$ up to an appropriate initial condition.
Naturally, the initial condition should be related to the regularization problem of the vacuum
of the Fock representation.
To see this, let us look at the unmodified representation $\rho_N$
on the ``vacuum state $\vert \varnothing)$,'' $\varnothing = (0, \ldots, 0)$,
\begin{align}
  (\varnothing \vert \Delta^{N-1}(K^+(z)) \vert \varnothing)
  &= \prod_{k=1}^{N} \tilde{\psi}(q_1^{-1}q_2^{k-1}v/z)
  = \prod_{k=1}^{N} \psi(q_1^{-1}q_2^{k-1}v/z) \psi(q_2^{k-1} q_3 z/v)^{-1} \CR
  &= \frac{1-q_2^N v/z}{1-q_2^N q_3 v/z} \frac{1-q_3 v/z}{1-v/z}. \label{inf}
\end{align}
This expression makes no sense in the limit of $N \to \infty$, but we can formally regularize it by specifying
the ordering of infinite products,
\begin{align}
  \psi(q_3v/z)^{-1} \prod_{k\geq 1} \left( \psi(q_1^{-1}q_2^{k-1}v/z) \psi( q_2^k q_3 v/z)^{-1} \right)
  = \psi(q_3v/z)^{-1}.
\end{align}
Therefore, in the modified representation $\bar{\rho}_N$,
one expects the vacuum expectation value to be
\begin{align}
  (\varnothing \vert K^+_N(z) \vert \varnothing)
  &= \prod_{k=1}^{N} \tilde{\psi}(q_1^{-1}q_2^{k-1}v/z) \beta^+_N(v/z) \CR
  &= \psi(q_3v/z)^{-1} = \gq^{-1} \frac{1-q_3 v/z}{1-v/z}, \quad \forall N \in \mathbb{N}
 .\label{reg1}
\end{align}
Thus, the problematic factor $\frac{1- q_2^N v/z}{1- q_2^N q_3 v/z}$ in \eqref{inf} has been
replaced by the factor $\gq^{-1}$ by the regularization.
Now \eqref{reg1} leads to the initial condition
\begin{align}
  \beta^+_1(v/z) = \psi(q_1^{-1}v/z)^{-1}.
\end{align}
Hence, the above prescription for the regularization gives
\begin{align}
  \beta_N^+(v/z) = \psi(q_1^{-1}q_2^{N-1}v/z)^{-1}
  = \psi(q_2^{-N} z/v) = \beta_N^-(z/v). \label{beta}
\end{align}
As concerns the action of $F(z)$, we also need some modification factor.
This factor should be precisely the same as $\beta_N^+(v/z)$ for the sake of \eqref{EFcoef}.
In fact, it satisfies the condition \eqref{idea} due to the vanishing property,
\begin{align}
  \beta^+_{N+1} F(z) [q_2^{N}v]_{-1}
  = \psi(q_1^{-1}q_2^N v/z)^{-1} \delta(q_1^{-1} q_2^{N}v/z) [q_2^{N}v]_{-2}
  =0.
\end{align}

Now we can write down the action of the infinite tensor product
representation $(\rho, \mathcal{F}(v))$ explicitly (see \eqref{DNK}--\eqref{DNF}),
\begin{align}
  K^+(z) \vert \lambda)
  &= \prod_{s=1}^{\ell(\lambda)} \tilde{\psi}(x_s v/z) ~\beta^+_{\ell(\lambda)}(v/z) \vert \lambda)
  = \prod_{s=1}^{\ell(\lambda)} \psi(x_s v/z) \prod_{s=1}^{\ell(\lambda)+1} \psi(q_2^{-1} x_s v/z)^{-1} \vert \lambda), \label{FockKp} \\
  K^-(z) \vert \lambda)
  &= \prod_{s=1}^{\ell(\lambda)} \tilde{\psi}(q_1^{-1}x_s^{-1} z/v) ~\beta^-_{\ell(\lambda)}(z/v) \vert \lambda)
  = \prod_{s=1}^{\ell(\lambda)} \psi(q_3 x_s^{-1} z/v)^{-1} \prod_{s=1}^{\ell(\lambda)+1} \psi(q_1^{-1} x_s^{-1} z/v) \vert \lambda), \label{FockKm} \\
  E(z) \vert \lambda)
  &= (1 -q_2) \sum_{j=1}^{\ell(\lambda)+1} \prod_{s=1}^{j-1} \tilde{\psi}(q_1^{-1}x_s/x_j) \delta(q_1 x_j v/z) \vert \lambda+1_j) \CR
  &= (1 -q_2)  \sum_{j=1}^{\ell(\lambda)+1} \prod_{s=1}^{j-1} \psi(q_1^{-1}x_s/x_j) \psi(q_3 x_s/x_j)^{-1} \delta(q_1x_j v/z) \vert \lambda+1_j), \label{FockE} \\
  F(z) \vert \lambda)
  &= (1 -q_2^{-1}) \sum_{j=1}^{\ell(\lambda)} \prod_{s=j+1}^{\ell(\lambda)} \tilde{\psi}(x_s/x_j) \delta(x_j v/z) ~\beta^+_{\ell(\lambda)}(1/x_j) \vert \lambda-1_j) \CR
  &=  (1 -q_2^{-1})  \sum_{j=1}^{\ell(\lambda)} \prod_{s=j+1}^{\ell(\lambda)} \psi(x_s/x_j) \prod_{s=j+1}^{\ell(\lambda)+1} \psi(q_2^{-1}x_s/x_j)^{-1} \delta(x_j v/z) \vert \lambda-1_j), \label{FockF}
\end{align}
where we have introduced the coordinates $x_s=x_{s,\lambda_s}$, $x_{i,j}=q_1^{j-1}q_2^{i-1}$.
As already mentioned, we can find the invariant subspace $\mathcal{F}(v)$
which consists of only partitions by investigating the positions of zeros appearing in the action of the creation operator $E(z)$
and the annihilation operator $F(z)$ \cite{FFJMM}.
This irreducible subrepresentation generated by the empty Young diagram $\varnothing$ is called Fock representation.
The Fock representation is the highest weight representation with the empty Young diagram $\varnothing$ being the highest weight state.
The generating function of eigenvalues of the vacuum is $(\varnothing\vert K^+(z) \vert \varnothing)=\psi(q_3v/z)^{-1}$.
Note that, similarly to the vector representation, the finite tensor representations have trivial level $(1,1)$,
however, now the Fock representation has the nontrivial level $(1,\gq)$ due to the chosen regularization \eqref{reg1}.

Thanks to the relation
\begin{align}
  G(x_{i,j}v/z) = \tilde{\psi}(x_{i,j}v/z) \tilde{\psi}(q_1^{-1}x_{i,j}v/z)^{-1},
\end{align}
one can rewrite the above formula in a symmetric way, for example,
\begin{align}
  (\lambda \vert K^+(z) \vert \lambda)
  = \psi_\varnothing(v/z) \prod_{(i,j)\in \lambda} G(x_{i,j} v/z),
\end{align}
where $\psi_\varnothing(v/z)$ is
\begin{align}
  \psi_\varnothing(v/z)
  = (\varnothing \vert K^+(z) \vert \varnothing)
  = \psi(q_3 v/z)^{-1}.
\end{align}

\subsubsection{MacMahon representation}
We want to define the MacMahon representation
as an appropriate limit of $(\pi_N, \mathcal{M}_N(v))$ similarly to the definition of the Fock representation \cite{FJMM},
\begin{align}
  \bigotimes_{k=1}^{N} \mathcal{F}(q_3^{k-1}v) \supset \mathcal{M}_N(v)
  \ni \vert \Lambda) =\bigotimes_{k=1}^{N} \vert \Lambda^{(k)}).
\end{align}
The notation of the plane partition ($3$d partition) $\Lambda$ is as follows
\begin{align}
  \Lambda = (\Lambda^{(1)}, \ldots , \Lambda^{(N)}), \quad \Lambda^{(k)} : \mbox{(2d) partition such that} \ \Lambda_i^{(k+1)} \leq \Lambda_i^{(k)}, \  (\forall i,k),
\end{align}
and we also use the notations
\begin{align}
  h(\Lambda) &= \min \{ k \geq 0 | (1,1,k+1) \not\in \Lambda \}, \\
  (i,j,k) \in \Lambda &\Leftrightarrow \Lambda_i^{(k)} \geq j.
\end{align}
We interpret as $\vert \Lambda^{(k)}) \in \mathcal{F}(q_3^{k-1}v)$ unless otherwise mentioned.

Let us now focus on the action of $K^\pm(z)$ and the regularization problem.
We denote as $K_N^\pm(z) = \bar{\pi}_N(K^\pm(z)) = \Delta^{N-1}(K^\pm(z)) \times \gamma_N^\pm$,
where $\gamma_N^\pm = \gamma_N^\pm((v/z)^\pm)$ is a modification factor which satisfies
$\gamma_N^+(v/z) = \gamma_N^-(z/v)$ as a rational function.
Since we have, for $\vert \varnothing) \in \mathcal{F}(q_3^N v)$,
\begin{align}
  (\varnothing \vert K^+(z) \vert \varnothing)\vert_{\mathcal{F}(q_3^N v)}
  = \psi(q_3^{N+1} v/z)^{-1},
\end{align}
we get a recursion relation for the modification factor $\gamma_N^+$ as
\begin{align}
  1=&\frac{(\Lambda \vert K_{N+1}^+(z) \vert \Lambda)\vert_{\mathcal{M}_{N+1}(v)}}{(\Lambda \vert K_{N}^+(z) \vert \Lambda)\vert_{\mathcal{M}_{N}(v)}}
  = \frac{\gamma^+_{N+1}}{\gamma^+_N} ~\psi(q_3^{N+1} v/z)^{-1},
  \end{align}
which means
\begin{align}
  \gamma^+_N
  = \psi(q_3^{N} v/z) \gamma^+_{N-1}
  = \gq^{N-1} \frac{1-q_3v/z}{1-q_3^N v/z} \gamma^+_1.
\end{align}
Hence, we can determines $\gamma^\pm_N$ up to an appropriate initial condition.
As concerns the initial condition, let us look at the unmodified representation $\pi_N$
on the vacuum state $\vert \varnothing)\in\mathcal{M}_{N}(v)$, $\varnothing=(\varnothing^{(1)}, \ldots, \varnothing^{(N)})$,
\begin{align}
  (\mathbf{\varnothing} \vert \Delta^{N-1}(K^+(z)) \vert \mathbf{\varnothing})\vert_{\mathcal{M}_{N}(v)}
  = \prod_{k=1}^{N} \psi(q_3^k v/z)^{-1}
  = \gq^{-N} \prod_{k=1}^{N} \frac{1-q_3^k v/z}{1-q_3^{k-1}v/z}
  = \gq^{-N} \frac{1-q_3^N v/z}{1-v/z}.
\end{align}
This expression makes no sense in the limit of $N \to \infty$,
but we cannot use the same strategy as in the previous subsection due to the presence of
the monomial factor $\gq^{-N}$. Here we have to regularize it in another way:
we formally substitute $\gq^N$ by an arbitrary parameter $K^{1/2}$.
Then, in the modified representation $\bar{\pi}_N$, we have
\begin{align}
  (\mathbf{\varnothing} \vert K_N^+(z) \vert \mathbf{\varnothing})\vert_{\mathcal{M}_{N}(v)}
  &= \prod_{k=1}^{N} \psi(q_3^k v/z)^{-1} ~\gamma^+_N(v/z)
  = K^{-1/2} \frac{1-K v/z}{1-v/z}, \label{reg2}
\end{align}
which leads to the initial condition
\begin{align}
  \gamma^+_1(v/z) = \frac{K^{-1/2}(1-K v/z)}{\gq^{-1}(1-q_3 v/z)}.
\end{align}
Hence, our regularization gives
\begin{align}
  \gamma^+_N(v/z) = \frac{K^{-1/2}(1-K v/z)}{\gq^{-N}(1-q_3^N v/z)}. \label{gamma}
\end{align}

Now we can write down the action of the MacMahon representation $(\pi, \mathcal{M}(K; v))$,
\begin{align}
  K^\pm(z) \vert \Lambda)
  &= \prod_{k=1}^{h(\Lambda)} (\Lambda^{(k)} \vert K^\pm(z) \vert \Lambda^{(k)})\vert_{\mathcal{F}(q_3^{k-1} v)} ~\gamma^\pm_{h(\Lambda)}((v/z)^{\pm1}) \vert \Lambda) \CR
  &= {\boldsymbol \psi}^\pm_{\varnothing}\left( K^{1/2}; (v/z)^{\pm1} \right) \prod_{(i,j,k)\in \Lambda} G((x_{ijk}v/z)^{\pm1})^{\pm1} \vert \Lambda), \label{MacK} \\
  E(z) \vert \Lambda)
  &= \sum_{k=1}^{h(\Lambda)+1} \sum_{i=1}^{\ell(\Lambda^{(k)})+1} \prod_{s=1}^{k-1} (\Lambda^{(s)} \vert K^-(q_1 x_{ijk}v) \vert \Lambda^{(s)})\vert_{\mathcal{F}(q_3^{s-1} v)} \CR
  & \qquad \times (\Lambda^{(k)}+1_i \vert E(z) \vert \Lambda^{(k)})\vert_{\mathcal{F}(q_3^{k-1} v)} \vert \Lambda+1^{(k)}_i), \label{MacE} \\
  F(z) \vert \Lambda)
  &= \sum_{k=1}^{h(\Lambda)} \sum_{i=1}^{\ell(\Lambda^{(k)})} (\Lambda^{(k)}-1_i \vert F(z) \vert \Lambda^{(k)})\vert_{\mathcal{F}(q_3^{k-1} v)} \CR
  & \qquad \times \prod_{s=k+1}^{h(\Lambda)} (\Lambda^{(s)} \vert K^+(x_{ijk}v) \vert \Lambda^{(s)})\vert_{\mathcal{F}(q_3^{s-1} v)} ~\gamma^+_{h(\Lambda)}(x_{ijk}^{-1}) \vert \Lambda-1^{(k)}_i), \label{MacF}
\end{align}
where we have introduced the coordinates $x_{ijk} = q_1^{j-1}  q_2^{i-1} q_3^{k-1} = q_1^j  q_2^i q_3^k $ and
the generating function of eigenvalues of the vacuum:
\begin{equation}
  {\boldsymbol \psi}^\pm_{\varnothing}(K^{1/2}; u)
 =  K^{\mp 1/2} \frac{1-K^{\pm} u}{1-u}.
\end{equation}
We note that $j=\Lambda_i^{(k)}$ is understood at the RHS of \eqref{MacE} and \eqref{MacF}.
The action of $K^\pm(z)$ has a manifestly symmetric expression
with respect to the permutation of $q_{1,2,3}$ but $E(z), F(z)$ do not.
In \cite{FJMM}, it was proved that the actions \eqref{MacK}--\eqref{MacF} indeed keep the subspace spanned by plane partitions.
This is assured by the $q_3$-shift of spectral parameters among adjacent Fock representations in the tensor product.
One can check that the factor
$(\Lambda^{(s)} \vert K^-(q_1 x_{ijk}v) \vert \Lambda^{(s)})\vert_{\mathcal{F}(q_3^{s-1} v)}$ in \eqref{MacE}
vanishes for $s=k-1$ when $\Lambda_i^{(k-1)} = \Lambda_i^{(k)}$,
which means that the action of $E(z)$ cannot add the box at $(i, \Lambda_i^{(k)}+1, k)$.
On the other hand, the factor $(\Lambda^{(s)} \vert K^+(x_{ijk}v) \vert \Lambda^{(s)})\vert_{\mathcal{F}(q_3^{s-1} v)}$
in \eqref{MacF} vanishes for $s=k+1$ when $\Lambda_i^{(k)} = \Lambda_i^{(k+1)}$,
which implies that the action of $F(z)$ cannot remove the box at $(i, \Lambda_i^{(k)}, k)$.
Thus the invariant subspace $\mathcal{M}(K; v)$ is generated by acting $E(z)$ on the empty plane partition, which is the highest weight state,
while the action of $F(z)$ keeps $\mathcal{M}(K; v)$.
In fact, there is a formula called shell formula \cite{FJMM, Prochazka:2015deb}, which expresses
the RHS of \eqref{MacE} and \eqref{MacF} as a finite sum over the concave and the convex corners of $\Lambda$.
Note that $\pi_N$ has level $(1,\gq^N)$ as an $N$-fold Fock representation,
however, now the MacMahon representation has level $(1,K^{1/2})$ with a continuous parameter $K$
due to the regularization \eqref{reg2}.

\subsection{Horizontal vertex operator representation}
\label{sec:hor}

When $C=\gq$, the Heisenberg part satisfies
\begin{align}
  [H_r, H_s]
  &= \delta_{r+s,0} \frac{[r]^2}{r} \gq^r(1-q_1^r)(1-q_2^r),
\end{align}
and we define the fundamental vertex operator as follows,
\begin{align}
  V^{\pm}(z) = \exp \left( \mp \sum_{r=1}^\infty \frac{H_{\pm r}}{[r]} z^{\mp r} \right).
\end{align}
The fundamental OPE by the normal ordering is
\begin{align}
  V^+(z) V^-(w) = s(w/z) : V^+(z) V^-(w) :,
\end{align}
where the scattering factor $s(z)$ is
\begin{align}
  s(z)
  = \frac{(1-\gq z)(1-\gq^{-1} z)}{(1-\gd z)(1-\gd^{-1} z)},
\end{align}
and satisfies the following formulas as a rational function,
\begin{align}
  s(z) = s(z^{-1}), \quad  G(z) = s(\gq^{-1} z)s(\gq z)^{-1} = G(z^{-1})^{-1}. \label{scat}
\end{align}

We can define the vertex operator representation of level $(\gq, 1)$
by making use of the fundamental vertex operator as follows:
\begin{align}
  E(z) \to \eta(z) &= V^-( \gq^{-\frac{1}{2}} z ) V^+(\gq^{\frac{1}{2}} z), \\
  F(z) \to \xi(z) &= V^-( \gq^{\frac{1}{2}} z )^{-1} V^+(\gq^{-\frac{1}{2}} z)^{-1}, \\
  K^\pm(\gq^{1/2}z) \to \varphi^{\pm}(z) &= V^{\pm}(\gq^{\pm1} z) V^{\pm}(\gq^{\mp 1} z)^{-1},
\end{align}
where the vacuum state $\vert 0 \rangle$ of the Fock space for the horizontal representation is defined by
the annihilation condition
\begin{align}
  H_r \vert 0 \rangle = 0, \quad r>0.
\end{align}
The shift of the argument in $K^\pm(z)$ is conventional.
Furthermore, for any $\gamma_2 \in \mathbb{C}^\times$,
we can employ a more general level $(\gq, \gamma_2)$ representation
with zero modes $\mathbf{e}(z), \mathbf{f}(z), \mathbf{k}^\pm(z)$,
\begin{align}
  E(z) &\to \eta(z) ~\mathbf{e}(z), \\
  F(z) &\to \xi(z) ~\mathbf{f}(z), \\
  K^{\pm}(\gq^{1/2} z) &\to \varphi^{\pm}(z) ~\mathbf{k}^\pm(z),
\end{align}
where the constant part of $\mathbf{k}^\pm(z)$ is fixed by the second level $\gamma_2$.
We denote the horizontal representation with these zero modes by
$\mathcal{H}=\mathcal{H}(\mathbf{k}^\pm(z), \mathbf{e}(z), \mathbf{f}(z))$.
The zero modes must satisfy appropriate relations.
In fact, we have
\begin{align}
    \mathbf{e}(z) \mathbf{f}(\gq^{\mp 1} z) = \mathbf{k}^\pm(\gq^{\mp 1/2}z) \label{EFzero}
\end{align}
from \eqref{EFcoef}, so $\mathbf{e}(z)$ can be determined from $\mathbf{f}(z)$ and $\mathbf{k}^{\pm}(z)$ uniquely
\begin{align}
  \mathbf{e}(z)
  = \mathbf{k}^\pm(\gq^{\mp 1/2}z) / \mathbf{f}(\gq^{\mp 1} z)
  = \sqrt{\mathstrut \mathbf{k}^+(\gq^{-1/2}z) \mathbf{k}^-(\gq^{1/2}z) /\mathbf{f}(\gq^{-1} z) \mathbf{f}(\gq z)}.
\end{align}
Furthermore, we make an ansatz that $\mathbf{k}^\pm(z)$ is independent of $z$:
\begin{align}
  \mathbf{k}^\pm(z) = \mathbf{k}^\pm(0) = \gamma_2^{\mp1}, \label{Kzero}
\end{align}
so that we can lift the modification factors \eqref{beta}, \eqref{gamma} of vertical representations to vertex operators uniquely.
Under the ansatz \eqref{Kzero}, there is a one parameter family of constant solutions to \eqref{EFzero}.
However, as we will see in section 4, one cannot keep $\mathbf{e}(z)$ and $\mathbf{f}(z)$ constant
for the existence of the intertwiner.
For example, the level $(\gq,\gq^N)$ representation used in \cite{Awata:2011ce} is
defined by
\begin{align}
\mathbf{k}^\pm(z) = \gq^{\mp N}, \qquad
\mathbf{e}(z) =  (\gq/z)^N u, \label{NE} \qquad
\mathbf{f}(z) = (\gq/z)^{-N} u^{-1},
\end{align}
where $u$ is the spectral parameter of the representation.
With the notation introduced above, we can express this horizontal representation
as $\mathcal{F}_u^{(\gq,\gq^{N})} = \mathcal{H}(\gq^{\mp N}, (\gq/z)^N u, (\gq/z)^{-N} u^{-1})$.

It is also useful to introduce the dual vertex operator $\tilde{V}^\pm(z)$ that satisfies
\begin{align}
 & V^+(z) \tilde{V}^-(w) = (1-w/z)^{-1} ~:V^+(z) \tilde{V}^-(w):, \\
  & \tilde{V}^+(z) V^-(w) = (1-w/z) ~:\tilde{V}^+(z) V^-(w):.
\end{align}
It is expressed explicitly as
\begin{align}
  \tilde{V}^\pm(z) = \exp\left( \mp\sum_{r=1}^{\infty} \Lambda_{\pm r}z^{\mp r} \right), \qquad
    \Lambda_r := \frac{\gq-\gq^{-1}}{k_r}H_r, \label{dualvertex}
\end{align}
where
\begin{align}
  k_r &= \prod_{i=1}^{3} (q_i^{r/2}-q_i^{-r/2}) = \prod_{i=1}^{3} (q_i^{r}-1) = \sum_{i=1}^{3} (q_i^{r}-q_i^{-r})
\end{align}
and we have
\begin{align}
  [\Lambda_r, H_s] = \delta_{r+s,0} \frac{[r]}{r}.
\end{align}


\section{DIM $R$-matrix for MacMahon representations}
\label{sec:dim-r-matrix}
In this section, we calculate the universal DIM $R$-matrix in the basis
of generalized triple Macdonald polynomials. It will be diagonal and
depend on the ratio of spectral parameters $x = \frac{u}{v}$, two
central charges $K_1$, $K_2$ and a pair of plane partitions $\Pi$,
$\Lambda$.

We use the formula for the diagonal (on the vertical representations)
part of the universal $R$-matrix from \cite{FJMM2}:
\begin{equation}
  \label{eq:2}
  \mathcal{R}_0 = (K \otimes 1 )^{\frac{1}{2} (1 \otimes d_1)} (1
  \otimes K )^{\frac{1}{2}  (d_1 \otimes 1)}
  \exp \left\{  - \sum_{n \geq 1} n k_n
    (h_{-n} \otimes h_n) \right\},
\end{equation}
where $K$ is the second component of the central charge vector, $d_1$ is the first grading operator
and $h_n$ are modes of the $K^{\pm}(z)$ currents defined as follows:
\begin{equation}
  \label{eq:3}
  K^{\pm}(z) = K^{\mp \frac{1}{2}} \exp \left\{ \sum_{n \geq 1} k_n h_{\pm n} z^{\mp n}  \right\}.
\end{equation}
For the vertical MacMahon representation, we have
\begin{align}
  \label{eq:4}
&  K^{+}(z)|\Pi, u) \CR
 & =   {\boldsymbol \psi}^{+}_{\varnothing}(K^{1/2}; \frac{u}{z})
    \prod_{(i,j,k)\in \Pi} \frac{\left(1
      - q_1^{i-1} q_2^j q_3^k \frac{u}{z} \right)\left(1 -
      q_1^i q_2^{j-1} q_3^k \frac{u}{z} \right)\left(1 -
      q_1^i q_2^j q_3^{k-1} \frac{u}{z} \right)}{\left(1 -
      q_1^{i+1} q_2^j q_3^k \frac{u}{z} \right)\left(1 -
      q_1^i q_2^{j+1} q_3^k \frac{u}{z} \right)\left(1
      - q_1^i q_2^j q_3^{k+1} \frac{u}{z} \right)} |\Pi, u),\\
 & K^{-}(z)|\Pi, u) \CR
&  =    {\boldsymbol \psi}^{-}_{\varnothing}(K^{1/2}; \frac{z}{u})
    \prod_{(i,j,k)\in \Pi} \frac{\left(1
      - q_1^{1-i} q_2^{-j} q_3^{-k} \frac{z}{u} \right)\left(1 - q_1^{-i}
      q_2^{1-j} q_3^{-k} \frac{z}{u} \right)\left(1 - q_1^{-i} q_2^{-j}
      q_3^{1-k} \frac{z}{u} \right)}{\left(1 - q_1^{-i-1} q_2^{-j} q_3^{-k}
      \frac{z}{u} \right)\left(1 - q_1^{-i} q_2^{-j-1} q_3^{-k} \frac{z}{u}
    \right)\left(1 - q_1^{-i} q_2^{-j} q_3^{-k-1} \frac{z}{u} \right)} |\Pi, u),
\end{align}
and therefore
\begin{equation}
  h_n|\Pi, u) = \frac{u^n}{|n|} \left[  \frac{1 -
      K^n}{k_n} + \sum_{(i,j,k)\in \Pi}
    q_1^{ni} q_2^{n j} q_3^{nk} \right]|\Pi, u)~.
\end{equation}

The $R$-matrix in two MacMahon representations is given by (we divide
by the vacuum matrix element to get rid of the overall scalar factor,
which can be evaluated separately)
\begin{multline}
  \label{eq:5}
  R_{\Pi\Lambda}^{K_1, K_2} \left( \frac{u}{v} \right) =\frac{( \Pi,u,K_1 | \otimes ( \Lambda, v, K_2|
    \mathcal{R}_0 |\Pi,u, K_1 ) \otimes |\Lambda, v,
    K_2)}{( \varnothing,u, K_1 | \otimes (
    \varnothing, v, K_2| \mathcal{R}_0 |\varnothing,u, K_1
   ) \otimes |\varnothing, v, K_2)}=\\
  = K_1^{\frac{|\Lambda|}{2}} K_2^{\frac{|\Pi|}{2}} \exp \Biggl\{ \sum_{n \geq 1} \left( \frac{u}{v} \right)^n
  \frac{1}{n} \Bigl[ (1 - K_1^{-n})\sum_{(i,j,k)\in \Lambda} q_1^{ni}
  q_2^{n j} q_3^{nk} - (1 - K_2^n) \sum_{(i,j,k)\in \Pi} q_1^{-ni}
  q_2^{-n j} q_3^{-nk} +\\
  - k_n \sum_{(i,j,k)\in \Pi} q_1^{-ni}
  q_2^{-n j} q_3^{-nk} \sum_{(a,b,c)\in \Lambda} q_1^{na} q_2^{n b}
  q_3^{nc}\Bigr] \Biggr\}=\\
  K_1^{\frac{|\Lambda|}{2}} K_2^{\frac{|\Pi|}{2}} \prod_{(i,j,k)\in
    \Pi} \prod_{(a,b,c)\in \Lambda} G^{-1} \left( \frac{u}{v} q_1^{a-i}
    q_2^{b-j} q_3^{c-k} \right) \prod_{(i,j,k)\in
    \Pi} \frac{1 - K_2 \frac{u}{v} q_1^{-i} q_2^{-j} q_3^{-k}}{1 -
    \frac{u}{v} q_1^{-i} q_2^{-j} q_3^{-k}} \prod_{(a,b,c) \in
    \Lambda} \frac{1 -  \frac{u}{v} q_1^a q_2^b q_3^c}{1 -
    K_1^{-1} \frac{u}{v} q_1^a q_2^b q_3^c},
\end{multline}
where $G(x) = \frac{(1- q_1^{-1} x)(1- q_2^{-1} x)(1- q_3^{-1} x)}{(1-
  q_1 x)(1- q_2 x)(1- q_3 x)}$.  The result is a rational function,
e.g.
\begin{equation}
  \label{eq:7}
  R_{[[1]]\varnothing}^{K_1, K_2} (x) = K_2^{\frac{1}{2}} \frac{1-x}{1
  - K_2 x},\qquad R_{\varnothing[[1]]}^{K_1, K_2} (x) =
K_1^{-\frac{1}{2}} \frac{K_1 - x}{1 - x}.
\end{equation}
It reduces to the $R$-matrix for the Fock representations in the limit of
$K_{1,2} \to q_3 = \frac{1}{q_1 q_2}$.

Notice that the $R$-matrix obeys a simple inversion identity
\begin{equation}
  \label{eq:1}
  R^{K_2,K_1}_{\Lambda \Pi} (x) = \frac{1}{R^{K_1,K_2}_{\Pi \Lambda} (x^{-1})}\ ,
\end{equation}
which does not allow one to use such an $R$-matrix for generating new knot invariants \cite[sect.12]{Awata:2016bdm}.
It also respects the symmetry of the DIM algebra under arbitrary
permutation of parameters $q_i$, which needs to be accompanied by the
corresponding transposition of the plane partitions $\Pi$ and
$\Lambda$, e.g.\ exchange of $q_i$ and $q_j$ corresponds to the
transposition in the $(i,j)$-plane.

In what follows, we build intertwining operators of the Fock
representations with the MacMahon ones and verify that the same $R$-matrix
determines their commutation relations.

\section{Intertwining operator for MacMahon representation}
\label{sec:int}

The intertwiner of the quantum toroidal algebra $U_{\gq,\gd}(\widehat{\widehat{\mathfrak{gl}}}_1)$
is the vertex operator that intertwines the tensor product of vertical and horizontal representations
with an appropriate horizontal representation.
Graphically, it is represented by a trivalent vertex with two \lq\lq horizontal\rq\rq\ edges and a single vertical edge.
As we explained earlier, the center of $U_{\gq,\gd}(\widehat{\widehat{\mathfrak{gl}}}_1)$ is two dimensional, and
one defines two levels $(\gamma_1, \gamma_2)$
for each representation.
In the following, we fix the horizontal representation as the vertex operator representations of free deformed bosons.
It has the unit first level $\gamma_1=\gq$.
On the other hand, we call representations with $\gamma_1=1$ vertical representation\footnote{
In the case of affine algebra, the evaluation representation is an example of the vertical representation.},
and there are various choices for the vertical representations, as we saw in the previous section.
Note that the condition $\gamma_1=1$ is kept intact under taking the tensor product of two
vertical representations.

In general, one can define the trivalent intertwiner $\Psi : \mathcal{V} \otimes \mathcal{H} \to \mathcal{H}'$
for a vertical representation $\mathcal{V}$ with a pair of horizontal representations $(\mathcal{H}, \mathcal{H}')$
by the intertwining condition
\begin{align}
  a \Psi = \Psi \Delta (a), \qquad
  \forall a \in U_{\gq,\gd}(\widehat{\widehat{\mathfrak{gl}}}_1).
\end{align}
Because $\mathcal{V}$ has a basis $\{ \alpha\}$ that simultaneously diagonalizes $K^{\pm}(z)$,
one can define the $\alpha$-component of the intertwiner as an operator between the horizontal representations
\begin{align}
  \Psi_\alpha(\bullet) = \Psi(\alpha \otimes \bullet) \colon \mathcal{H} \to \mathcal{H}', \quad \bullet \in \mathcal{H}.
\end{align}
Since $C_1=1$ and $C_2 = \gq$ for the vertical and the horizontal representations, the definition of the coproduct
\eqref{cpE} -- \eqref{cpK-} implies the following intertwining relations for the $\alpha$-components:
\begin{align}
  K^{+}(z) \Psi_\alpha &= (\alpha \vert  K^{+}(z) \vert \alpha) ~\Psi_\alpha K^{+}(z), \\
  K^{-}(\gq z) \Psi_\alpha &= (\alpha \vert  K^{-}(z) \vert \alpha) ~\Psi_\alpha K^{-}(\gq z), \\
  E(z) \Psi_\alpha &= \sum_{\beta} (\beta \vert E(z) \vert \alpha) ~\Psi_\beta + (\alpha \vert K^{-}(z) \vert \alpha) ~\Psi_\alpha E(z), \\
  F(z) \Psi_\alpha &= \sum_{\beta} (\beta \vert F(\gq z) \vert \alpha) ~\Psi_\beta K^{+} (\gq z) + \Psi_\alpha F(z).
\end{align}
Note that, in the above formulas, operators at the right side of $\Psi_\alpha$
act on $\mathcal{H}$, while operators on the left side of $\Psi_\alpha$ act on $\mathcal{H}'$.
The matrix elements $(\beta \vert X \vert \alpha)$ are computed in $\mathcal{V}$.
Our tasks are to construct the component of the intertwiner in the vertex operator formalism (section \ref{sec:hor})
and to specify admissible horizontal representations.
Note that we can typically specify the relative conditions between $\mathcal{H}$ and $\mathcal{H}'$
but there seemingly remains some freedom for each of them.
At first, we will construct the vector intertwiner, which is essentially the same as the one given by \cite{FHHSY} before.
Then, as a composition of the vector intertwiner with an appropriate ordering,
we will obtain the Fock intertwiner firstly given by \cite{Awata:2011ce}.
Finally, using the same strategy we will construct the MacMahon intertwiner from the Fock intertwiner.
The $N$-fold Fock intertwiner was constructed in \cite{Bourgine:2017jsi},
but the spectral parameters of Fock representations were independent in their construction.
Our MacMahon intertwiner corresponds to an appropriate $N \to \infty$ limit of the intertwiner in \cite{Bourgine:2017jsi}
when the spectral parameters are correlated so that the infinite tensor product leads to the MacMahon representation.

In the following, we use the notation $\mathbf{x}(z)$ for the zero modes of the vertex operator
representation $\mathcal{H}$, while $\mathbf{x}'(z)$ for those of $\mathcal{H}'$,
where $\mathbf{x}=\mathbf{e}, \mathbf{f}, \mathbf{k}^{\pm}$.
We also denote the second level of $\mathcal{H}, \mathcal{H}'$ by $\gamma, \gamma'$ unless otherwise mentioned.

\subsection{Vector intertwiner}
\subsubsection{Definition of the vector intertwiner}

Let us consider the intertwining operator for the vertical vector representation $V(v)$.
We define the $n$-component of the vector intertwiner by
\begin{align}\label{vcomponent}
  \mathbb{I}_n(v)(\bullet) = \mathbb{I}([v]_{n-1} \otimes \bullet) \colon \mathcal{H} \to \mathcal{H}', \quad \bullet \in \mathcal{H},
\end{align}
where $\{ [v]_{n-1}\}_{n\in\mathbb{Z}}$ is the basis of $V(v)$.
Explicit intertwining relations for $\mathbb{I}_n(v)$ are
\begin{align}
  K^{+}(z) \mathbb{I}_{n}(v) &= \tilde{\psi}(q_1^{n-1}v/z) ~\mathbb{I}_{n}(v) K^{+}(z), \\
  K^{-}(\gq z) \mathbb{I}_{n}(v) &= \tilde{\psi}(q_1^{-n}z/v) ~\mathbb{I}_{n}(v) K^{-}(\gq z), \\
  E(z) \mathbb{I}_{n}(v) &= (1-q_2) \delta(q_1^nv/z) ~\mathbb{I}_{n+1}(v) + \tilde{\psi}(q_1^{-n}z/v) ~\mathbb{I}_{n}(v) E(z), \label{vintE} \\
  F(z) \mathbb{I}_{n}(v) &= (1-q_2^{-1}) \delta(\gq^{-1}q_1^{n-1}v/z) ~\mathbb{I}_{n-1}(v) K^{+} (\gq z) + \mathbb{I}_{n}(v) F(z). \label{vintF}
\end{align}
In the next subsection, we find that there are consistency conditions between the source $\mathcal{H}$ and the target $\mathcal{H}'$,
namely $\gamma'=\gamma$ on the second level and
$\mathbf{e}'(z)=q_2^{-1}\mathbf{e}(z)$, $\mathbf{f}'(z)=q_2\mathbf{f}(z)$ on the zero modes.

\subsubsection{Construction of the vector intertwiner}

We define the operator $\mathbb{I}_n(v)$ between two horizontal representations as
\begin{align}\label{vintertwin}
  \mathbb{I}_n(v)
  &= z_n \tilde{\mathbb{I}}_n(v), \quad
  \tilde{\mathbb{I}}_n(v) = \tilde{\mathbb{I}}_0(q_1^n v), \quad
  n \in \mathbb{Z}, \\
  \tilde{\mathbb{I}}_0(v)
  &= \exp\left( -\sum_{r=1}^{\infty}\frac{H_{-r}}{[r]} \frac{(\gq^{-1/2} v)^r}{1-q_1^r} \right)
  \exp\left( -\sum_{r=1}^{\infty}\frac{H_{r}}{[r]} \frac{(\gq^{1/2}q_1^{-1} v)^{-r}}{1-q_1^r} \right),
\end{align}
where $z_n=z_n(v)$ is a stack of zero modes
\begin{align}
  z_0 = 1, \quad
  z_n = q_2^{-n} \prod_{j=1}^{n} \mathbf{e}(q_1^{j-1}v) \quad (n>0), \quad
  z_n = q_2^{-n} \prod_{j=n}^{-1} \mathbf{e}(q_1^{j}v)^{-1} \quad (n<0).
\end{align}
Note that  $\tilde{\mathbb{I}}_n(v)$ is a lift of the function \eqref{psitilde} to the vertex operator, and we can formally interpret $\tilde{\mathbb{I}}_n(v)$ as an infinite product form
\begin{align}
  :\prod_{j=n+1}^{\infty} \eta(q_1^{j-1}v)^{-1}:.
\end{align}
Using the fundamental OPE relations
\begin{align}
  V^+(z) \mathbb{I}_0(v)
  = \frac{1-\gq^{1/2}q_2v/z}{1-\gq^{1/2}v/z} ~:V^+(z) \mathbb{I}_0(v):, \quad
  \mathbb{I}_0(v) V^-(z)
  = \frac{1-\gq^{1/2}q_1z/v}{1-\gq^{1/2}q_3^{-1}z/v} ~:\mathbb{I}_0(v) V^-(z):,
\end{align}
one can check the following relations for the zero-component $\mathbb{I}_0(v)$:
\begin{align}
  \varphi^+(\gq^{-1/2}z) \mathbb{I}_0(v)
  &= \tilde{\psi}(q_1^{-1} v/z) ~\mathbb{I}_0(v) \varphi^+(\gq^{-1/2}z), \label{vint1} \\
  \mathbb{I}_0(v) \varphi^-(\gq^{1/2}z)
  &= \tilde{\psi}(z/v)^{-1} ~\varphi^-(\gq^{1/2}z) \mathbb{I}_0(v), \label{vint2} \\
  \xi(z) \mathbb{I}_0(v)
  = \frac{1-\gq v/z}{1-\gq q_2v/z} ~:\xi(z) \mathbb{I}_0(v):, \quad
  &\mathbb{I}_0(v) \xi(z)
  = \frac{1-\gq q_3^{-1}z/v}{1-\gq q_1 z/v} ~:\mathbb{I}_0(v) \xi(z):, \\
  \eta(z) \mathbb{I}_0(v)
  = \frac{1-q_2 v/z}{1-v/z} ~:\eta(z) \mathbb{I}_0(v):, \quad
  &\mathbb{I}_0(v) \eta(z)
  = \frac{1-q_1z/v}{1-q_3^{-1}z/v} ~:\mathbb{I}_0(v) \eta(z):.
\end{align}
Actually, we have defined $\mathbb{I}_0(v)$ so that it satisfies the relations \eqref{vint1}, \eqref{vint2},
which imply $\gamma=\gamma'$ on the second level.
The relative relation for the zero modes is determined from the intertwining relations for $F(z)$ and $E(z)$.
To see it, let us check the intertwining relation for the $0$-component for $F(z)$.
To obtain the relation \eqref{vintF}, we find that the additional factor $q_2$ of $F(z)$ on the left is necessary,
compared to $F(z)$ on the right, namely
\begin{align}
  q_2\xi(z) \mathbf{f}(z) \mathbb{I}_0(v) -\mathbb{I}_0(v) \xi(z) \mathbf{f}(z)
  = (1-q_2^{-1}) \delta\left( \frac{q_1^{-1}v}{\gq z} \right) q_2 \mathbf{e}(q_1^{-1}v)^{-1} \tilde{\mathbb{I}}_{-1}(v) \varphi^+(\gq^{1/2}z) \gamma^{-1},
\end{align}
where we have used the relation \eqref{EFzero} and
\begin{align}
  \delta\left( \frac{q_1^{-1}v}{\gq z} \right) \xi(z)
  = \delta\left( \frac{q_1^{-1}v}{\gq z} \right) \eta(q_1^{-1}v)^{-1} \varphi^+(\gq^{1/2}z).
\end{align}
This is the reason why the vector intertwiner should shift the horizontal zero mode by $q_2^{-1}$.
In the same way, one can check
\begin{align}
  q_2^{-1}\eta(z) \mathbf{e}(z) \mathbb{I}_0(v) -\tilde{\psi}(z/v) \mathbb{I}_0(v) \eta(z) \mathbf{e}(z)
  = (1-q_2) \delta(v/z) q_2^{-1} \mathbf{e}(v) \tilde{\mathbb{I}}_{1}(v).
\end{align}
The intertwining relation for general components can be also checked easily
by making use of the relation among the functions $G(z), \tilde{\psi}(z)$ and $s(z)$.

If we choose the horizontal representation $\mathcal{H}=\mathcal{F}_u^{(\gq,\gq^N)}$ used in \cite{Awata:2011ce},
the zero mode sector is defined by \eqref{NE} and $\mathcal{H}'=\mathcal{F}_{q_2^{-1}u}^{(\gq,\gq^N)}$.
Then, the zero mode stack $z_n=z_n(N;u \vert v)$ is
\begin{align}
  z_n = (u/q_2)^n \prod_{j=1}^{n} \left( \frac{\gq}{q_1^{j-1} v} \right)^N \quad (n>0), \qquad
  z_n = (u/q_2)^n \prod_{j=n}^{-1} \left( \frac{\gq}{q_1^{j-1} v} \right)^N \quad (n<0).
\end{align}

Essentially the same vector intertwiner was given in appendix of \cite{FHHSY},
where they computed the generating function of the components of the intertwiner \eqref{vcomponent}.
By the simple $n$-dependence \eqref{vintertwin} of the components, it is easy to see
that the intertwiner in \cite{FHHSY} agrees with ours.

\subsection{Fock intertwiner}
\subsubsection{Definition of the Fock intertwiner}

Let us consider the intertwining operator for the vertical Fock representation $\mathcal{F}(v)$.
With the basis $\{ \vert\lambda)\}$ of $\mathcal{F}(v)$,
explicit intertwining relations for the $\lambda$-component $\Phi_\lambda(v)$ are
\begin{align}
  K^{+}(z) \Phi_\lambda(v) &= (\lambda \vert  K^{+}(z) \vert \lambda) ~\Phi_\lambda(v) K^{+}(z), \\
  K^{-}(\gq z) \Phi_\lambda(v) &= (\lambda \vert  K^{-}(z) \vert \lambda) ~\Phi_\lambda(v) K^{-}(\gq z), \\
  E(z) \Phi_\lambda(v) &= \sum_{k=1}^{\ell(\lambda)+1} (\lambda+1_k \vert E(z) \vert \lambda) ~\Phi_{\lambda+1_k}(v) + (\lambda \vert K^{-}(z) \vert \lambda) ~\Phi_\lambda(v) E(z), \\
  F(z) \Phi_\lambda(v) &= \sum_{k=1}^{\ell(\lambda)} (\lambda-1_k \vert F(\gq z) \vert \lambda) ~\Phi_{\lambda-1_k}(v) K^{+} (\gq z) + \Phi_\lambda(v) F(z),
\end{align}
where matrix elements can be read from \eqref{FockKp}--\eqref{FockF}.
In the next subsection, we obtain as consistency conditions the relative shift $\gamma'=\gq\gamma$
of the second level and
$\mathbf{e}'(z)=(-\gq v/z)\mathbf{e}(z)$, $\mathbf{f}'(z)=(-\gq v/z)^{-1}\mathbf{f}(z)$
on the relative shift of the zero modes.

\subsubsection{Construction of the Fock intertwiner}

We construct the Fock intertwiner in a parallel way to the construction of the Fock representation (section \ref{sec:Fockrep})
in contrast to a rather direct way of \cite{Awata:2011ce}.
In our approach, taking the tensor product of the vertical representations is
realized by merely composing the corresponding vector intertwiners.
And the modification factor $\beta_n^\pm(v/z)$ is lifted to the corresponding vertex operator $B_n(v)$.

We begin with specifying the modification operator $B_n(v)$.
This is realized by making use of the dual vertex operator \eqref{dualvertex} as
\begin{align}
  B_n(v)
  &= \tilde{V}^-(\gq^{1/2} q_2^nv) \tilde{V}^+(\gq^{3/2} q_2^n v)^{-1},
\end{align}
and satisfies the following relations
\begin{align}
  \varphi^+(\gq^{-1/2}z) B_n(v)
  &= \gq \beta^+_n(v/z) ~B_n(v) \varphi^+(\gq^{-1/2}z), \label{beta1} \\
  B_n(v) \varphi^-(\gq^{1/2}z)
  &= \gq \beta^-_n(z/v)^{-1} ~\varphi^-(\gq^{1/2}z) B_n(v), \label{beta2} \\
  \left( \frac{\gq}{z}(-q_2^{n}v) \right)^{-1} \xi(z) B_n(v) &- B_n(v) \xi(z)
  = 0 , \label{beta3} \\
  \frac{\gq}{z}(-q_2^{n}v)\eta(z) B_n(v) &- \beta^-_n(z/v) B_n(v) \eta(z)
  = -\gq \delta(q_2^{n} v/z) :\eta(z) B_n(v):. \label{beta4}
\end{align}
Note that we can formally interpret $B_n(v)$ as an infinite product form
\begin{align}
  :\prod_{i,j=1}^{\infty} \eta(q_1^{j-1}q_2^{n+i-1}v)^{-1}:.
\end{align}
\eqref{beta1} and \eqref{beta2} mean that the operator $B_n(v)$ exactly corresponds
to the modification factor $\beta_n^\pm(v/z)$ up to the monomial factor $\gq$, and this discrepancy leads to the relative level shift.
Furthermore, \eqref{beta3} and \eqref{beta4} lead to the relative shift of the zero modes.
We can summarize the conditions for $B_n(v) \colon \mathcal{H} \to \mathcal{H}''$ as follows
\begin{align}
  \gamma'' = \gq \gamma, \quad
  \mathbf{e}''(z) = (-\gq q_2^nv/z) \mathbf{e}(z), \quad
  \mathbf{f}''(z) = (-\gq q_2^nv/z)^{-1} \mathbf{f}(z).
\end{align}

Now we can define the operator $\Phi_{\lambda}(v)$ between two horizontal representations
as the following composition\footnote{One should be careful of the ordering of constituent vector intertwiners.}
of vector intertwiners with the modification operator $B_n(v)$
\begin{align}
  \Phi_{\lambda}(v)
  &= z_\lambda \tilde{\Phi}_\lambda(v) :
  \mathcal{H} \to \mathcal{H}'' \to \mathcal{H}', \\
  \tilde{\Phi}_\lambda(v)
  &= \mathcal{G}^{[n]} \cdot \tilde{\mathbb{I}}^{[n]}_\lambda(v) B_n(v),
  \quad n > \ell(\lambda), \label{indep1}
\end{align}
where $\tilde{\mathbb{I}}^{[n]}_\lambda(v)
 = \tilde{\mathbb{I}}_{\lambda_1}(v) \circ \cdots \circ \tilde{\mathbb{I}}_{\lambda_n}(q_2^{n-1}v)$
and the coefficient $\mathcal{G}^{[n]}$ is defined by the normal ordering\footnote{We have defined as $\lambda_n=0$ for $n>\ell(\lambda)$.}
\begin{align}
  \tilde{\mathbb{I}}^{[n]}_{\varnothing}(v) B_n(v)
  = (\mathcal{G}^{[n]})^{-1} :\tilde{\mathbb{I}}^{[n]}_{\varnothing}(v) B_n(v):.
\end{align}
Finally, $z_\lambda=z_\lambda(v)$ is a stack of zero modes (see the discussion of the intertwiner zero modes in Appendix C)
\begin{align}
  z_\lambda(v)
  = \prod_{i=1}^{\ell(\lambda)} \prod_{j=1}^{\lambda_i} \left( -\gq q_2^{i-1} x_{i,j}^{-1} \right) \mathbf{e}(x_{i,j}v)
  = q_2^{n(\lambda)} (-\gq)^{|\lambda|} \prod_{(i,j)\in\lambda} x_{i,j}^{-1} \mathbf{e}(x_{i,j}v).
\end{align}
Note that the definition \eqref{indep1} is independent of $n>\ell(\lambda)$ thanks to the relation
\begin{align}
  &:\tilde{\mathbb{I}}^{[n]}_\lambda(v) B_n(v): = :\tilde{\mathbb{I}}^{[n+1]}_\lambda(v) B_{n+1}(v): \CR
  &= \exp\left( \sum_{r=1}^{\infty} \frac{H_{-r}}{[r]}(\gq^{-1/2}v)^r\left( \sum_{(i,j)\in\lambda}x_{i,j}^r - \frac{1}{(1-q_1^r)(1-q_2^r)} \right) \right) \CR
  & \quad \times \exp\left( -\sum_{r=1}^{\infty} \frac{H_{r}}{[r]}(\gq^{1/2}v)^{-r}
  \left( \sum_{(i,j)\in\lambda}x_{i,j}^{-r} - \frac{q_3^{-r}}{(1-q_1^r)(1-q_2^r)} \right) \right),
\end{align}
which can be also understood from the formal infinite product form of them.
The consistency conditions between the horizontal representations in $\Phi_\lambda(v)\colon \mathcal{H} \to \mathcal{H}'$
can be read as
\begin{align}
  \gamma' = \gq \gamma, \quad
  \mathbf{e}'(z) = (-\gq v/z) \mathbf{e}(z), \quad
  \mathbf{f}'(z) = (-\gq v/z)^{-1} \mathbf{f}(z).
\end{align}
Indeed, these zero modes of $\mathcal{H}'$ satisfy the relations \eqref{EFzero}.
We can confirm that the operator $\Phi_\lambda(v)$ agrees with the AFS intertwiner \cite{Awata:2011ce},
and the factor $\mathcal{G}_\lambda$, which is defined independently of $n>\ell(\lambda)$ by
\begin{align}
  \tilde{\mathbb{I}}_\lambda^{[n]}(v) B_n(v)
  = (\mathcal{G}_\lambda)^{-1} (\mathcal{G}^{[n]})^{-1} :\tilde{\mathbb{I}}_\lambda^{[n]}(v) B_n(v):,
\end{align}
plays the same role as $c_\lambda$ in \cite{Awata:2011ce}.
An explicit form of $\mathcal{G}_\lambda$ is
\begin{align}
  \mathcal{G}_\lambda
  &= \prod_{\square \in \lambda} \left( 1-q_1^{-a_\lambda(\square)}q_2^{l_\lambda(\square)+1} \right)
  = q_1^{-n(\lambda')} q_2^{n(\lambda)+|\lambda|} c_\lambda, \\
  c_\lambda
  &= \prod_{\square \in \lambda} ( 1-q_1^{a_\lambda(\square)}q_2^{-l_\lambda(\square)-1} ).
\end{align}
If we choose the AFS type horizontal representation $\mathcal{H}=\mathcal{F}_u^{(\gq,\gq^N)}$, where the zero mode action is defined by \eqref{NE},
then $\mathcal{H}'=\mathcal{F}_{-vu}^{(\gq,\gq^{N+1})}$, and the zero mode stack $z_\lambda=z_\lambda(N;u\vert v)$ is
\begin{align}
  z_\lambda(N;u\vert v)
  &= \prod_{i=1}^{\ell(\lambda)} z_{\lambda_i}(N+1;-q_2^i vu\vert q_2^{i-1}v)
  = \prod_{i=1}^{\ell(\lambda)} (-q_2^{i-1}vu)^{\lambda_i} \prod_{j=1}^{\lambda_i} \left( \frac{\gq}{q_1^{j-1}q_2^{i-1}v} \right)^{N+1} \CR
  &= q_2^{n(\lambda)} (-vu)^{|\lambda|} \prod_{(i,j)\in \lambda} \left( \gq x_{i,j}^{-1} v^{-1} \right)^{N+1},
\end{align}
where $n(\lambda) = \displaystyle{\sum_{j=1}^{\ell(\lambda)}} (j-1)\lambda_j$ and
$x_{i,j} = q_1^{j-1}q_2^{i-1}$.

One can check the Fock intertwining relations by making use of the vector intertwining relation.
For example, to check the intertwining relation with $E(z)$, one can compute as follows:
\begin{align}
&  E(z) \Phi_\lambda(v)  - \prod_{s=1}^{n} \tilde{\psi}(q_1^{-\lambda_s}q_2^{1-s}z/v) ~\beta^-_{n}(z/v) ~\Phi_\lambda(v) E(z)  \CR
& = z_\lambda \mathcal{G}^{[n]} \sum_{k=1}^{n} (-\gq q_2^k v/z) \prod_{s=1}^{k-1} \tilde{\psi}(q_1^{-\lambda_s}q_2^{1-s}z/v)
 \tilde{\mathbb{I}}_{\lambda_1}(v) \cdots  \tilde{\mathbb{I}}_{\lambda_{k-1}}( q_2^{k-2}v )  \CR
 & ~~~~~~~~~\times   [[ E(z),   \tilde{\mathbb{I}}_{\lambda_{k}}( q_2^{k-1}v ) ]]_k
 \tilde{\mathbb{I}}_{\lambda_{k+1}}( q_2^{k} v) \cdots  \tilde{\mathbb{I}}_{\lambda_{n}}( q_2^{n-1}v ) B_{n}(v) \CR
 &~~~+  z_\lambda \mathcal{G}^{[n]} \prod_{s=1}^{n} \tilde{\psi}(q_1^{-\lambda_s}q_2^{1-s}z/v) \tilde{\mathbb{I}}_{\lambda}^{[n]} (v)
 \left( E(z) B_{n}(v) - \beta^{-} (z/v) B_{n}(v)E(z)   \right),
\end{align}
where we have introduced the \lq\lq weighted\rq\rq\ commutator by
\begin{align}
 [[ A, B ]]_k  = AB -  \tilde{\psi}(q_1^{-\lambda_k}q_2^{1-k}z/v) BA\ ,
\end{align}
and the zero mode ${\bf e}(z)$ of $E(z)$ should be adjusted appropriately according to the space on which $E(z)$ acts.
First of all, the last term on the RHS vanishes due to the factor $\tilde{\psi}(q_2^{-n+1} z/v)\delta(q_2^n v/z)$.
The intertwining relation for the vector intertwiner tells us that
\begin{align}
  [[ E(z),   \tilde{\mathbb{I}}_{\lambda_{k}}( q_2^{k-1}v ) ]]_k
= (1- q_2) \delta (q_1 x_k v/z) q_2^{-1} \mathbf{e}(z) \tilde{\mathbb{I}}_{\lambda_{k}+1}(q_2^{k-1}v).
\end{align}
Hence, we obtain
\begin{align}
&  E(z) \Phi_\lambda(v)  - ( \lambda \vert K^{-}(z) \vert \lambda ) ~\Phi_\lambda(v) E(z)  \CR
& = (1 - q_2) z_\lambda \mathcal{G}^{[n]}\cdot q_2^{-1} \sum_{k=1}^{n} (-\gq q_2^k v/z) \prod_{s=1}^{k-1} \tilde{\psi}(q_1^{-\lambda_s}q_2^{1-s}z/v)
 \tilde{\mathbb{I}}_{\lambda_1}(v) \cdots  \tilde{\mathbb{I}}_{\lambda_{k-1}}( q_2^{k-2}v )  \CR
 & ~~~~~~~~~\times  \delta (q_1 x_k v/z) \mathbf{e}(z)  \tilde{\mathbb{I}}_{\lambda_{k}+1}(q_2^{k-1}v)
 \tilde{\mathbb{I}}_{\lambda_{k+1}}( q_2^{k} v) \cdots  \tilde{\mathbb{I}}_{\lambda_{n}}( q_2^{n-1}v )  B_{n}(v)  \CR
& =  \sum_{k=1}^{n}  ( \lambda_k +1 \vert E(z) \vert \lambda) \Phi_{\lambda + 1_k} (v).
 \end{align}
As concerns the intertwining relation with $F(z)$, one can compute as follows:
\begin{align}
&  F(z) \Phi_\lambda(v)  - \Phi_\lambda(v) F(z)
= z_\lambda \mathcal{G}^{[n]} \sum_{k=1}^{n} (-\gq q_2^k v/z)^{-1}
\tilde{\mathbb{I}}_{\lambda_1}(v) \cdots  \tilde{\mathbb{I}}_{\lambda_{k-1}}( q_2^{k-2}v ) \CR
 & ~~\times [F(z), \tilde{\mathbb{I}}_{\lambda_{k}}( q_2^{k-1}v ) ]  ~\tilde{\mathbb{I}}_{\lambda_{k+1}}( q_2^{k} v) \cdots
\tilde{\mathbb{I}}_{\lambda_{n}}( q_2^{n-1}v ) B_{n}(v)
+  z_\lambda \mathcal{G}^{[n]} \cdot \tilde{\mathbb{I}}_{\lambda}^{[n]} (v) [F(z), B_{n}(v)].
\end{align}
Note that this time we do not have to use the \lq\lq weighted\rq\rq\ commutator.
Again, the last term vanishes and, by substituting the intertwining relations for the vector intertwiner, we obtain
\begin{align}
&  F(z) \Phi_\lambda(v)  - \Phi_\lambda(v) F(z)  \CR
& = (1 - q_2^{-1}) z_\lambda \mathcal{G}^{[n]}\cdot  q_2 \sum_{k=1}^{n} (-\gq q_2^k v/z)^{-1} \tilde{\mathbb{I}}_{\lambda_1}(v) \cdots
\tilde{\mathbb{I}}_{\lambda_{k-1}}( q_2^{k-2}v ) \CR
& ~~~~~~~~\times \delta (\gq^{-1}x_k v/z) \mathbf{f}(z) \tilde{\mathbb{I}}_{\lambda_k -1}( q_2^{k-1}v ) K^+(\gq z)
~\tilde{\mathbb{I}}_{\lambda_{k+1}}( q_2^{k} v) \cdots  \tilde{\mathbb{I}}_{\lambda_{n}}( q_2^{n-1}v ) B_{n}(v) \CR
& = (1 - q_2^{-1}) z_\lambda \mathcal{G}^{[n]} \cdot q_2 \sum_{k=1}^{n} (-\gq q_2^k v/z)^{-1} \prod_{s=k+1}^{n} \tilde{\psi}(q_1^{\lambda_s -1}q_2^{s-1}v/z) \beta_n^+(v/z)
  \tilde{\mathbb{I}}_{\lambda_1}(v) \cdots  \tilde{\mathbb{I}}_{\lambda_{k-1}}( q_2^{k-2}v ) \CR
& ~~~~~~~~\times \delta (\gq^{-1}x_k v/z) \mathbf{f}(z) \tilde{\mathbb{I}}_{\lambda_k -1}( q_2^{k-1}v )
~\tilde{\mathbb{I}}_{\lambda_{k+1}}( q_2^{k} v) \cdots  \tilde{\mathbb{I}}_{\lambda_{n}}( q_2^{n-1}v ) B_{n}(v) ~K^+(\gq z) \CR
& =  \sum_{k=1}^{n}  ( \lambda_k -1 \vert F(z) \vert \lambda) \Phi_{\lambda - 1_k} (v) K^+(\gq z).
\end{align}

\subsection{MacMahon intertwiner}
\subsubsection{Definition of the MacMahon intertwiner}

Let us consider the intertwining operator for the vertical MacMahon representation $\mathcal{M}(K;v)$.
We define the MacMahon intertwiner $\Xi(K;v)$ by the following condition
\begin{align}
  \Xi(K;v) : \mathcal{M}(K;v) \otimes \mathcal{H} \to \mathcal{H}', \quad
  a \Psi = \Psi \Delta (a), \quad
  a \in U_{\gq,\gd}(\widehat{\widehat{\mathfrak{gl}}}_1)\ ,
\end{align}
and the $\Lambda$-component of the MacMahon intertwiner by
\begin{align}
  \Xi_\Lambda(\bullet) = \Xi(\vert\Lambda) \otimes \bullet) \colon \mathcal{H} \to \mathcal{H}', \quad \bullet \in \mathcal{H},
\end{align}
where $\{ \vert\Lambda)\}$ is the basis of $\mathcal{M}(K;v)$.
Explicit intertwining relations for $\Xi_\Lambda(K;v)$ are
\begin{align}
  K^{+}(z) \Xi_\Lambda(v) &= (\Lambda \vert  K^{+}(z) \vert \Lambda) ~\Xi_{\Lambda}(v) K^{+}(z), \\
  K^{-}(\gq z) \Xi_\Lambda(v) &= (\Lambda \vert  K^{-}(z) \vert \Lambda) ~\Xi_{\Lambda}(v) K^{-}(\gq z), \\
  E(z) \Xi_\Lambda(v) &= \sum (\Lambda + 1_i^{(k)} \vert E(z) \vert \Lambda) ~\Xi_{\Lambda + 1_i^{(k)}}(v) + (\Lambda \vert K^{-}(z) \vert \Lambda) ~\Xi_{\Lambda}(v) E(z), \\
  F(z) \Xi_\Lambda(v) &= \sum (\Lambda - 1_i^{(k)} \vert F(\gq z) \vert \Lambda ) ~\Xi_{\Lambda - 1_i^{(k)}}(v) K^{+} (\gq z) + \Xi_\Lambda (v) F(z),
\end{align}
where the matrix elements can be read from \eqref{MacK}--\eqref{MacF}.
In the next subsection, we specify the constraint on the relations among the zero modes.

\subsubsection{Construction of the MacMahon intertwiner}

We construct the MacMahon intertwiner in the same way as the construction of the Fock intertwiner in section 4.2.
We begin with specifying the vertex operator $\Gamma_n(K;v)$ that produces
the modification factor $\gamma_n^\pm(z/v)$ given by \eqref{gamma} in the OPE relation.
This operator is realized by
\begin{align}
  \Gamma_n(K;v)
  = \exp\left( \sum_{r=1}^{\infty}\frac{H_{-r}}{[r]} \frac{q_3^{nr}-K^r}{k_r} (\gq^{-1/2} v)^r \right)
  \exp\left( \sum_{r=1}^{\infty}\frac{H_{r}}{[r]} \frac{q_3^{-nr}-K^{-r}}{k_{r}} (\gq^{1/2} v)^{-r} \right),
\end{align}
and satisfies the following fundamental relations
\begin{align}
  V^+(z) \Gamma_n(K;v)
  &= \exp\left( \sum_{r=1}^{\infty}\frac{1}{r} \frac{q_3^{nr}-K^r}{1-q_3^r} (\gq^{1/2} v/z)^r \right) :V^+(z) \Gamma_n(K;v):, \\
  \Gamma_n(K;v) V^-(z)
  &= \exp\left( -\sum_{r=1}^{\infty}\frac{1}{r} \frac{q_3^{-nr}-K^{-r}}{1-q_3^{r}} (\gq^{3/2} v/z)^{-r} \right) :\Gamma_n(K;v) V^-(z):.
\end{align}
Note that we can formally interpret $\Gamma_n(K;v)$ as an infinite product form as
\begin{align}
  :\prod_{i,j,k=1}^{\infty} \eta(q_1^{j-1}q_2^{i-1} q_3^{n+k-1}v)^{-1} \eta(q_1^{j-1} q_2^{i-1} q_3^{k-1}Kv):.
\end{align}
Then we find the following relations:
\begin{align}
  \varphi^+(\gq^{-1/2}z) \Gamma_n(K;v)
  &= \frac{K^{1/2}}{\gq^n} \gamma^+_n(v/z) ~\Gamma_n(K;v) \varphi^+(\gq^{-1/2}z), \label{gamma1}\\
  \Gamma_n(K;v) \varphi^-(\gq^{1/2}z)
  &= \frac{K^{1/2}}{\gq^n} \gamma^-_n(z/v)^{-1} ~\varphi^-(\gq^{1/2}z) \Gamma_n(K;v) \label{gamma2},
\end{align}
which mean that the operator $\Gamma_n(K;v)$ produces
the modification factor $\gamma_n^\pm(v/z)$ up to a monomial factor.
This discrepancy corresponds to a shift of the second level:
\begin{align}
  \gamma'' = \gq^{-n} K^{1/2} \gamma.
\end{align}
Furthermore, the normal orderings with $\xi$ and $\eta$ are evaluated to give
\begin{align}
  \xi(z) \Gamma_n(K;v)
  &= \frac{ (\gq q_3^n v/z; q_3)_{\infty}}{ (K \gq v/z; q_3)_{\infty} } :\xi(z) \Gamma_n(K;v):, \\
  \Gamma_n(K;v) \xi(z)
  &= \frac{(K^{-1} \gq z/v; q_3)_{\infty}}{ (\gq q_3^{-n} z/v; q_3)_{\infty}}  :\xi(z) \Gamma_n(K;v):, \\
  \eta(z) \Gamma_n(K;v)
  &= \frac{(K v/z; q_3)_{\infty}} { (q_3^n v/z; q_3)_{\infty}}  :\eta(z) \Gamma_n(K;v):, \\
  \Gamma_n(K;v) \eta(z)
  &= \frac{ (q_3^{-n} z/v; q_3)_{\infty}} {(K^{-1} z/v; q_3)_{\infty}} :\eta(z) \Gamma_n(K;v):.
\end{align}
In order to obtain the relation $F(z) \Gamma_n(K;v)-\Gamma_n(K;v)F(z)=0$, we need the relative shift
\begin{align}
  \left[ \frac{(\gq q_3^n v/z; q_3)_{\infty}}{(K \gq v/z; q_3)_{\infty}} \right]^{-1}
  \frac{(K^{-1} \gq z/v; q_3)_{\infty}}{(\gq q_3^{-n} z/v; q_3)_{\infty}}
  = \frac{\theta_{q_3}(\gq K v/z)}{\theta_{q_3}(\gq q_3^n v/z)}, \label{Mzeroshift}
\end{align}
where the theta function $\theta_{q_3}(z)$ is defined by \eqref{theta}.
Note that this in turn determines the relative shift for $\mathbf{e}(z)$.
As a result, we have
\begin{align}
  \frac{\theta_{q_3}(\gq K v/z)}{\theta_{q_3}(\gq q_3^n v/z)} ~\xi(z) \Gamma_n(K;v) - \Gamma_n(K;v) \xi(z)
  &= 0, \label{gamma3} \\
  \frac{K^{1/2}}{\gq^{n}} \frac{\theta_{q_3}(q_3^n v/z)}{\theta_{q_3}(K v/z)} ~\eta(z) \Gamma_n(K;v) - \gamma_n^-(z/v) ~\Gamma_n(K;v) \eta(z)
  &=0. \label{gamma4}
\end{align}
Note that \eqref{gamma4} vanishes by itself,
while the corresponding equation \eqref{beta4} vanishes only after multiplying with the factor $\tilde{\psi}(q_2^{-n+1}z/v)$.
We can summarize the relative condition for $\Gamma_n(K;v) \colon \mathcal{H} \to \mathcal{H}''$ as follows
\begin{align}
  \mathbf{e}''(z) = \frac{K^{1/2}}{\gq^{n}} \frac{\theta_{q_3}(q_3^n v/z)}{\theta_{q_3}(K v/z)} \mathbf{e}(z), \qquad
  \mathbf{f}''(z) = \frac{\theta_{q_3}(\gq K v/z)}{\theta_{q_3}(\gq q_3^n v/z)} \mathbf{f}(z).
\end{align}

Now we can define the operator $\Xi_{\Lambda}(K;v)$ between two horizontal representations
as a well-ordered stack of the Fock intertwiners with the vertex operator $\Gamma_n(K;v)$
\begin{align}
  \Xi_{\Lambda}(K;v)
  &= z_\Lambda \tilde{\Xi}_\Lambda(K;v) : \mathcal{H}\to\mathcal{H}''\to\mathcal{H}', \\
  \tilde{\Xi}_\Lambda(K;v)
  &= \mathcal{M}^{[n]}(K) \tilde{\Phi}^{[n]}_\Lambda(v) \Gamma_n(K;v),
  \quad n > h(\Lambda), \label{indep2}
\end{align}
where the factor $\mathcal{M}^{[n]}(K)$ is defined as\footnote{We have defined as $\Lambda^{(n)}= \varnothing$ for $n>h(\Lambda)$.}
\begin{align}
  \tilde{\Phi}^{[n]}_{\varnothing}(v) \Gamma_n(K;v)
  &= \mathcal{M}^{[n]}(K)^{-1} :\tilde{\Phi}^{[n]}_{\varnothing}(v) \Gamma_n(K;v):, \\
  \tilde{\Phi}^{[n]}_\Lambda(v)
  &= \tilde{\Phi}_{\Lambda^{(1)}}(v) \circ \cdots \circ \tilde{\Phi}_{\Lambda^{(n)}}(q_3^{n-1}v)\ ,
\end{align}
and $z_\Lambda=z_\Lambda(K;v)$ is the contribution from the zero modes,
\begin{align}
  z_\Lambda(K;v)
  = \prod_{k=1}^{h(\Lambda)} \prod_{(i,j)\in\Lambda^{(k)}}
  \frac{K^{1/2}}{\gq^{k-1}} \frac{\theta_{q_3}(q_3^{k-1}/x_{ijk})}{\theta_{q_3}(K/x_{ijk})} \mathbf{e}(x_{ijk}v).
\end{align}
Note that the definition \eqref{indep2} is independent of $n>h(\Lambda)$ thanks to the relation
\begin{align}
  &:\tilde{\Phi}^{[n]}_\Lambda(v) \Gamma_n(K;v): = :\tilde{\Phi}^{[n+1]}_\Lambda(v) \Gamma_{n+1}(K;v): \\
  &= \exp\left( \sum_{r=1}^{\infty} \frac{H_{-r}}{[r]}(\gq^{-1/2}v)^r\left( \sum_{(i,j,k)\in\Lambda}x_{ijk}^r + \frac{1-K^r}{k_r} \right) \right) \CR
  & \quad \times \exp\left( -\sum_{r=1}^{\infty} \frac{H_{r}}{[r]}(\gq^{1/2}v)^{-r}\left( \sum_{(i,j,k)\in\Lambda}x_{ijk}^{-r} - \frac{1-K^{-r}}{k_r} \right) \right),
\end{align}
which can be also understood from the formal infinite product form of them.
The relative condition of horizontal representations for $\Xi_\Lambda(K;v)\colon \mathcal{H} \to \mathcal{H}'$ can be read as
\begin{align}
  \gamma' = K^{1/2} \gamma, \quad
  \mathbf{e}'(z) = K^{1/2}\frac{\theta_{q_3}(v/z)}{\theta_{q_3}(Kv/z)} \mathbf{e}(z), \quad
  \mathbf{f}'(z) = \frac{\theta_{q_3}(\gq Kv/z)}{\theta_{q_3}(\gq v/z)} \mathbf{f}(z),
\end{align}
where we have used the formula in Appendix \ref{sec:formula}.
Indeed, these zero modes of $\mathcal{H}'$ satisfy the relations \eqref{EFzero}
as long as those of $\mathcal{H}$ satisfy the relations \eqref{EFzero}.
Hence, the operator $\Xi_\Lambda(K;v)$ exists for each horizontal representation $\mathcal{H}$.

The intertwining relations for $\Xi_{\Lambda}(K;v)$ essentially follow from those for the Fock intertwiners.
A proof can be done almost in parallel with the computations we have shown in the check of
the intertwining relations for the Fock intertwiners from those of the vector intertwiners.


\section{MacMahon $R$-matrix from the commutation of intertwiners}
\label{sec:r-matrix-from}
Following the technique developed for the Fock representations, we can
evaluate the commutator of two intertwining operators
$\Xi_{\Pi}(K;z)$, just constructed. For the moment, we omit the zero
modes. The result is
\begin{multline}
  \label{eq:9}
  \Xi_{\Pi}(K_1;z_1) \Xi_{\Lambda}(K_2;z_2) =
  \Xi_{\Lambda}(K_2;z_2) \Xi_{\Pi}(K_1;z_1) \frac{\Upsilon(K_1,
    K_2|z_1, z_2)}{R^{K_1,
      K_2}_{\Pi \Lambda}\left( \frac{z_1}{z_2} \right)}\times\\
  \times
  K_1^{-\frac{|\Lambda|}{2}} K_2^{-\frac{|\Pi|}{2}} \prod_{(i,j,k) \in
    \Pi} \frac{\theta_{q_3} \left( K_2 \frac{z_2}{z_1} q_1^{-i} q_2^{-j}
      q_3^{-k} \right)}{\theta_{q_3} \left( \frac{z_2}{z_1} q_1^{-i} q_2^{-j}
      q_3^{-k} \right)} \prod_{(a,b,c) \in
    \Lambda} \frac{\theta_{q_3} \left( \frac{z_2}{z_1} q_1^a q_2^b
      q_3^c \right)}{\theta_{q_3} \left( K_1^{-1}  \frac{z_2}{z_1} q_1^a q_2^b
      q_3^c \right)},
\end{multline}
where
\begin{multline}
  \label{eq:10}
  \Upsilon_{q_1, q_2, q_3}(K_1, K_2| z_1, z_2) = \exp \Biggl[ \sum_{n \geq 1}
    \frac{1}{n} \frac{1}{(1-q_1^n)(1-q_2^n)(1-q_3^n)^2} \Bigl( \left(
      \frac{z_2}{z_1} \right)^n (1 - K_1^{-n})(1 - K_2^n)-\\
    - \left(
        \frac{z_1}{z_2} \right)^n (1 - K_1^n) (1 - K_2^{-n}) \Bigr) \Biggr]\ ,
\end{multline}
and $\theta_q(x) = \prod_{k \geq 0} (1 - q^{k+1}) (1 - q^k x) (1 - q^{k+1}
x^{-1})$. The extra theta-functions in Eq.~\eqref{eq:9} are in fact
precisely cancelled by the zero modes of the intertwiners. Notice that
they depend on the \emph{states} $\Pi$ or $\Lambda$ separately (though
the dependence on the \emph{both} spectral parameters is nontrivial).

We conclude that the commutation relations for the intertwiners that
we have constructed in sec.~\ref{sec:int} indeed feature the MacMahon
$R$-matrix from sec.~\ref{sec:dim-r-matrix}. The whole picture of the
intertwiners and $R$-matrices which we have presented in this paper is
therefore consistent.

\section{Conclusions}
\label{sec:conclusions}
In this paper, we have introduced novel intertwining operators for the DIM
algebra. These operators feature the MacMahon representations
(representations on plane partitions), whose role in the DIM algebra is analogous
to the role of free field representations for (quantum) affine Lie
algebras. Our intertwining operators generalize the refined topological
vertices \cite{IKV,GIKV,AK0,Taki,AK}: they depend on a pair of ordinary Young diagrams and on one
$3d$ Young diagram, and on an extra parameter: the central charge
associated with the MacMahon representation. For quantized values of
the central charge, our vertices reproduce the refined topological
vertices, or a ``strip'' combination thereof.

We also write down explicitly the DIM $R$-matrix acting on the tensor
product of MacMahon representations (assuming their central charges
are aligned along the preferred direction) and prove that it
determines the commutation relations of the intertwiners.

The next logical step on our way is to build the generalization of the
network formalism \cite{MMZ, AKMMMOZ} with some legs having the MacMahon
representations and investigate the resulting partition functions and corresponding constraint algebras
\cite{qq1,qq2,NZ,MMZ, AKMMMOZ}. It is also very interesting to understand if there is a corresponding
Type IIB string construction. Since a pair of central charges of the DIM
representation corresponds to the $(p,q)$-charges of the 5-brane in
Type IIB, it is not immediately clear what would the MacMahon representation
with an \emph{arbitrary} (non-integer) central charge correspond to.

Using the technique of \cite{Awata:2017cnz, Awata:2017lqa},
one can derive the $(q,t)$-KZ equations for the DIM intertwiners containing not only Fock
representations, but the MacMahon ones too. However, in this approach, the
MacMahon representations are restricted to lie only on the ``vertical''
legs. It would be interesting to lift this restriction, but, to this end,
one needs an intertwiner of several MacMahon representations, which
have not yet been described.  In \cite{Awata:2016bdm}, we found that
the DIM $R$-matrix $ {\cal R}_{\lambda \mu}(x)$ for the Fock representations appears
in the difference equation
\begin{equation}
\label{Nek}
{N_{\lambda \mu}\left( \frac{q}{t} x \right)}
= \left(\frac{q}{t}\right)^{\frac{1}{2}(|\lambda|+|\mu|)}
   {\cal R}_{\lambda \mu}(x)^{-1}N_{\lambda \mu}(x) ,
\end{equation}
satisfied by the Nekrasov factor $N_{\lambda \mu}(x)$;
\begin{align}
 N_{\lambda \mu}(x) &= \prod_{(i,j)\in \lambda} \left( 1 - x
    q^{\lambda_i - j} t^{i - \mu^{\mathrm{T}}_j+1} \right)
  \prod_{(i,j)\in \mu} \left( 1 - x q^{-\mu_i + j-1} t^{-
      \lambda^{\mathrm{T}}_j+i} \right)\nonumber\\
  &= \exp \left[ - \sum_{n \geq 1}
    \frac{1-t^n}{n
      (1-q^n)}  x^n \sum_{i,j}
    (q^{n(\lambda_i - \mu_j)} -1) t^{n(j-i)} \right].
\end{align}
It is an interesting problem to solve a generalization of \eqref{Nek}
with ${\cal R}_{\lambda \mu}(x)$ replaced by the MacMahon
$R$-matrix $R_{\Pi \Lambda}^{K_1, K_2}(x)$.
The solution should be regarded as a generalized
Nekrasov factor $N_{ \Pi \Lambda}^{K_1, K_2}(x)$.
It is natural to expect that the factor is related to a norm of
the triple Macdonald functions \cite{Zenk}.

Another interesting aim might be the generalization of our story to
the case of MacMahon modules with nontrivial asymptotics, i.e.\ to
$3d$ Young diagrams with nontrivial ``ends" along the coordinate
axes \cite{FJMM, BFM}. We think one can use the same strategy by making a judicious
choice of the spectral parameters of constituent Fock representations.
Thus, the formulas should be quite similar to those presented above.

Finally, one can try to build the most general representation of the DIM algebra,
in which the \emph{both} central charges (not just one, as in the MacMahon
case) are completely arbitrary. To our knowledge, such representations
have never been studied, so it might lead to some unexpected surprises.

\subsection*{Acknowledgements}

Our work is supported in part by Grants-in-Aid for Scientific Research
(17K05275) (H.A.), (15H05738, 18K03274) (H.K.) and JSPS Bilateral Joint Projects (JSPS-RFBR collaboration)
``Topological Field Theories and String Theory: from Topological Recursion
to Quantum Toroidal Algebra'' from MEXT, Japan. It is also partly supported by the grant of the Foundation for the Advancement of Theoretical Physics ``BASIS" (A.Mir. and A.Mor.), by  RFBR grants 16-01-00291 (A.Mir.) and 16-02-01021 (A.Mor. and Y.Z.), by joint grants 17-51-50051-YaF, 18-51-05015-Arm-a (A.Mir., A.Mor. and Y.Z.), 18-51-45010-IND-a (A.Mir. and A.Mor.). The work of Y.Z. was supported in part by INFN and by the ERC Starting Grant 637844-HBQFTNCER.

\appendix

\section{Conventions and useful functions}

The intriguing triality of the quantum toroidal algebra $U_{\gq,\gd}(\widehat{\widehat{\mathfrak{gl}}}_1)$ becomes
manifest, when we use the parameters $(q_1,q_2,q_3)$ with $q_1 q_2 q_3 =1$.
These parameters are also natural from the point of view of
plane partitions used in the MacMahon representation.
In this paper, we use the following convention\footnote{The parameters $q_2$ and $q	_3$ are exchanged, compared with \cite{FFJMM} and \cite{FJMM}.}:
\begin{align}
  q_1 = \gd \gq^{-1} =q, \quad q_2 =  \gd^{-1} \gq^{-1} = t^{-1},  \quad q_3 = \gq^{2} = t/q.
\end{align}
The parameters $(q,t)$ can be identified with those appear in the Macdonald functions.
The formulas for the vertex operator representation can be expressed neatly by introducing the following factor:
\begin{align}
  k_r &= \prod_{i=1}^{3} (q_i^{r/2}-q_i^{-r/2}) = \prod_{i=1}^{3} (q_i^{r}-1) = \sum_{i=1}^{3} (q_i^{r}-q_i^{-r})
\end{align}
The function
\begin{align}
  G(z) &= \frac{(1- q_1^{-1}z)(1- q_2^{-1}z)(1- q_3^{-1}z)}{(1- q_1z)(1- q_2z)(1- q_3z)}, \qquad G(z) G(z^{-1}) =1.
\end{align}
derived from the structure function $g(z,w) = (z-q_1w)(z-q_2w)(z-q_3w)$
of $U_{\gq,\gd}(\widehat{\widehat{\mathfrak{gl}}}_1)$ appears in the formula of the MacMahon representation.
Starting from the fundamental rational function
\begin{align}
  \psi(z) = \gq \frac{1-q_3^{-1}z}{1-z} = \psi(q_3/z)^{-1},
\end{align}
we can reconstruct the function $G(x)$ as follows:
\begin{align}
\tilde{\psi}(z) = \psi(z) \psi(q_2^{-1}z)^{-1}, \qquad
G(z) = \tilde{\psi}(z) \tilde{\psi}(q_1^{-1} z)^{-1}.
\end{align}
These functions appear in the generating functions of the eigenvalues of $K^{\pm}(z)$ in the vertical representations.
On the other hand, in the vertex operator representation, the function $G(z)$ arise as
\begin{align}
G(z) = s(\gq^{-1} z)s(\gq z)^{-1}
\end{align}
from the scattering factor
\begin{align}
  s(z)
  = \frac{(1-\gq z)(1-\gq^{-1} z)}{(1-\gd z)(1-\gd^{-1} z)}
\end{align}
in the OPE relation of the vertex operators.

\section{Properties of $\theta$-function}
\label{sec:formula}

We naturally encounter the following $\theta$-function with parameter $p=q_3=\gq^2$
in the construction of the MacMahon intertwiner:
\begin{align}
  \theta_p(z) &:=  (p;p)_{\infty}(z;p)_{\infty} (pz^{-1};p)_{\infty}
  = (1-z) \prod_{k=1}^{\infty}(1-p^{k})(1-p^k z)(1-p^k z^{-1}), \label{theta} \\
  (z;p)_{\infty} &:= \prod_{k=0}^{\infty}(1-p^k z).
\end{align}
From the Jacobi triple product formula
\begin{align}
  \theta_p(z)
  = \sum_{n \in \mathbb{Z}} p^{\frac{n}{2}(n-1)}(-z)^n,
\end{align}
we see $\theta_p(z)$ is an elliptic function with quasi double periodicity.
In this paper, only ratios of $\theta$-functions appear in the formula,
for example see \eqref{Mzeroshift}, hence we can omit the factor $(p;p)_{\infty}$ in \eqref{theta} safely.
The function $\theta_p(z)$ satisfies the following relations
\begin{align}
  \theta_p(z^{-1}) &= -z^{-1} \theta_p(z), \\
  \theta_p(zp^n) &= (-z)^{-n} p^{-\frac{n}{2}(n-1)} \theta_p(z), \quad n \in \mathbb{Z}.
\end{align}


\section{Zero modes of the Fock intertwiner}
\label{sec:zero-modes}
In this section, we write down the zero modes of the Fock space
intertwining operators \cite{Awata:2011ce} in a more convenient notation.

The horizontal Fock representations are characterized by the integer
values of the two central charges: $K_1 =  q_3 = \frac{t}{q}$ and $K_2 =
\frac{t}{q}^N$, and by the complex spectral parameter $u$. The zero modes
of the Fock representation of the DIM algebra should therefore be
written as operators acting on the vectors $|N, u \rangle$. Let us
introduce an operator $P_1$, which by definition gives the eigenvalue $u$
when acting on $|N, u \rangle$:
\begin{equation}
  \label{eq:12}
  P_1 |N, u \rangle = u |N, u \rangle.
\end{equation}
Let us write the DIM generators $x^{\pm}(z)$, $\psi^{\pm}(z)$ from
\cite{Awata:2011ce} in terms of $P_1$, and $c_2 = \log_{q_3} K_2$:
\begin{align}
  \label{eq:13}
  x^{+}(z) &= P_1 \left( \sqrt{\frac{q}{t}} z \right)^{- c_2} \eta(z), \\
  x^{-}(z) &= P^{-1}_1 \left( \sqrt{\frac{q}{t}} z \right)^{c_2} \xi(z),\\
  \psi^{+}(z) &= \left( \frac{q}{t} \right)^{\frac{c_2}{2}} \varphi^{+}(z), \\
  \psi^{-}(z) &= \left( \frac{q}{t} \right)^{-\frac{c_2}{2}}
  \varphi^{-}(z),
\end{align}
where $\eta(z)$, $\xi(z)$ and $\varphi^{\pm}(z)$ denote the exponentials
of the free boson (non-zero) modes.

The intertwiner $\Psi(w)$ with the vertical Fock representation of
central charges $K_1 = 1$, $K_2 = \frac{t}{q}$ and the spectral parameter $w$
changes both the spectral parameter and the central charge of the
horizontal representation. To account for this, we introduce the
operators $Q_1$ and $e_2$ canonically conjugate to $P_1$, and $c_2$:
\begin{align}
  \label{eq:14}
  Q_1 |u,N\rangle &= u \frac{\partial}{\partial u}| u, N\rangle,\\
  e_2 |u,N\rangle &= |u,N+1\rangle.
\end{align}
The intertwiner is then given by
\begin{align}
  \label{eq:15}
  \Phi_{\lambda}(w) &= e_2 (-w)^{Q_1} \prod_{(i,j) \in \lambda} \left(
    (-w P_1) \left( w q^{j-\frac{1}{2}} t^{\frac{1}{2} - i}
    \right)^{-c_2 - 1}
  \right)  \widetilde{\Phi}_{\lambda}(w)
  \end{align}
where $\widetilde{\Phi}_{\lambda}(w)$ contains only nonzero bosonic
modes.

It is easy to derive the commutation relations of the intertwiners
with the generators, as well as between the intertwiners themselves
from the canonical commutation relations
\begin{align}
  \label{eq:17}
  Q_1 P_1 &= P_1 (Q_1 + 1),\\
  c_2 e_2 & = e_2 (c_2 + 1).
\end{align}

Notice that only part of the zero modes in
Eqs.~\eqref{eq:15} depends on the diagram living in the
vertical Fock module: the essential part (conjugate to $P_1$ and
$c_2$) is independent of the diagram.


\end{document}